\documentclass[aps,prl,amsmath,amssymb,
showpacs, showkeys,
superscriptaddress,
nofootinbib,
nobibnotes,
preprint, 
reprint, 
twocolumn,
]{revtex4-1}

\usepackage[latin9]{inputenc}
\usepackage{longtable}
\usepackage{textcomp}
\usepackage{amsmath}
\usepackage{amssymb}
\usepackage{graphicx}
\usepackage{epstopdf}
\usepackage[hidelinks]{hyperref}
\usepackage{epsfig}
\usepackage{dcolumn}
\usepackage{bm}

\newcolumntype{C}[1]{>{\centering\arraybackslash}p{#1}}

\makeatletter

\providecommand{\tabularnewline}{\\}

\date{\today}

\@ifundefined{textcolor}{}{%
 \definecolor{BLACK}{gray}{0}
 \definecolor{WHITE}{gray}{1}
 \definecolor{RED}{rgb}{1,0,0}
 \definecolor{GREEN}{rgb}{0,1,0}
 \definecolor{BLUE}{rgb}{0,0,1}
 \definecolor{CYAN}{cmyk}{1,0,0,0}
 \definecolor{MAGENTA}{cmyk}{0,1,0,0}
 \definecolor{YELLOW}{cmyk}{0,0,1,0}
}

\makeatother

\begin{document}

\title{Normal and intruder configurations in $^{34}$Si populated in the $\beta^-$ decay of $^{34}$Mg and $^{34}$Al}

%%AUTHOR LIST

%% MAIN AUTHORS 

\author{R.~Lic\u{a}} 
\affiliation{``Horia Hulubei" National Institute for Physics and Nuclear Engineering, RO-077125 Bucharest, Romania}
\affiliation{CERN, CH-1211 Geneva 23, Switzerland} 

\author{F.~Rotaru}   \affiliation{``Horia Hulubei" National Institute for Physics and Nuclear Engineering, RO-077125 Bucharest, Romania}	

\author{M.J.G.~Borge}   \affiliation{CERN, CH-1211 Geneva 23, Switzerland}											
\affiliation{Instituto de Estructura de la Materia, CSIC, Serrano 113 bis, E-28006 Madrid, Spain}
		
\author{S.~Gr\'{e}vy}	 \affiliation{GANIL, CEA/DRF-CNRS/IN2P3, Bvd Henri Becquerel, 14076 Caen, France}	
\affiliation{UMR 5797, CNRS/IN2P3, Universit\'{e} de Bordeaux, Chemin du Solarium, 33175 Gradignan Cedex, France}

\author{F.~Negoi\c{t}\u{a}}   \affiliation{``Horia Hulubei" National Institute for Physics and Nuclear Engineering, RO-077125 Bucharest, Romania}

\author{A.~Poves}	 \affiliation{Departamento de F\'{i}sica Te\'{o}rica, Universidad Aut\'{o}noma de Madrid, E-28049 Madrid, Spain}	
\affiliation{Instituto de  F\'{i}sica Te\'{o}rica, UAM-CSIC E-28049 Madrid, Spain}

\author{O.~Sorlin}	 \affiliation{GANIL, CEA/DRF-CNRS/IN2P3, Bvd Henri Becquerel, 14076 Caen, France}

%% ALPHABETICAL Co-authors

\author{A.N.~Andreyev}   \affiliation{University of York, Department of  Physics, York YO10 5DD, N Yorkshire, United Kingdom}	

\author{R.~Borcea}   \affiliation{``Horia Hulubei" National Institute for Physics and Nuclear Engineering, RO-077125 Bucharest, Romania}		

\author{C.~Costache}   \affiliation{``Horia Hulubei" National Institute for Physics and Nuclear Engineering, RO-077125 Bucharest, Romania}	

\author{H.~De~Witte}   \affiliation{KU Leuven, Instituut voor Kern- en Stralingsfysica, Celestijnenlaan 200D, 3001 Leuven, Belgium}		
	
\author{L.M.~Fraile}   \affiliation{Grupo de F\'isica Nuclear \& IPARCOS, Facultad de CC. F\'isicas, Universidad Complutense, CEI Moncloa, 28040 Madrid, Spain}											

\author{P.T.~Greenlees}   \affiliation{University of Jyvaskyla, Department of Physics, P.O. Box 35, FI-40014 University of Jyvaskyla, Finland}	
					\affiliation{Helsinki Institute of Physics, University of Helsinki, P.O. Box 64, FIN-00014 Helsinki, Finland }
			
\author{M.~Huyse}   \affiliation{KU Leuven, Instituut voor Kern- en Stralingsfysica, Celestijnenlaan 200D, 3001 Leuven, Belgium}	

\author{A.~Ionescu}   \affiliation{``Horia Hulubei" National Institute for Physics and Nuclear Engineering, RO-077125 Bucharest, Romania}
 \affiliation{University of Bucharest, Faculty of Physics, Atomistilor 405, Bucharest-Magurele, Romania}

\author{S.~Kisyov}   
\affiliation{Faculty of Physics, University of Sofia ``St. Kliment Ohridski", 1164 Sofia, Bulgaria}
\affiliation{``Horia Hulubei" National Institute for Physics and Nuclear Engineering, RO-077125 Bucharest, Romania}

\author{J.~Konki}   \affiliation{University of Jyvaskyla, Department of Physics, P.O. Box 35, FI-40014 University of Jyvaskyla, Finland}
           \affiliation{Helsinki Institute of Physics, University of Helsinki, P.O. Box 64, FIN-00014 Helsinki, Finland }
           \affiliation{CERN, CH-1211 Geneva 23, Switzerland}

\author{I.~Lazarus}   \affiliation{STFC Daresbury, Daresbury, Warrington WA4 4AD, UK}	

\author{M.~Madurga}   \affiliation{CERN, CH-1211 Geneva 23, Switzerland}												
\author{N.~M\u{a}rginean}   \affiliation{``Horia Hulubei" National Institute for Physics and Nuclear Engineering, RO-077125 Bucharest, Romania}			
						
\author{R.~M\u{a}rginean}   \affiliation{``Horia Hulubei" National Institute for Physics and Nuclear Engineering, RO-077125 Bucharest, Romania}				

\author{C.~Mihai}   \affiliation{``Horia Hulubei" National Institute for Physics and Nuclear Engineering, RO-077125 Bucharest, Romania}											

\author{R.~E.~Mihai}   \affiliation{``Horia Hulubei" National Institute for Physics and Nuclear Engineering, RO-077125 Bucharest, Romania}
		
\author{A.~Negret}   \affiliation{``Horia Hulubei" National Institute for Physics and Nuclear Engineering, RO-077125 Bucharest, Romania}											

\author{F.~Nowacki}   \affiliation{Universit\'e de Strasbourg, IPHC, 23 rue du Loess, F-67037 Strasbourg, France}											\affiliation{CNRS, UMR 7178, F-67037 Strasbourg, France}											

\author{R.D.~Page}   \affiliation{Department of Physics, Oliver Lodge Laboratory, University of Liverpool, Liverpool L69 7ZE, United Kingdom}	
							
\author{J.~Pakarinen}   \affiliation{University of Jyvaskyla, Department of Physics, P.O. Box 35, FI-40014 University of Jyvaskyla, Finland}											
                        \affiliation{Helsinki Institute of Physics, University of Helsinki, P.O. Box 64, FIN-00014 Helsinki, Finland }

\author{V.~Pucknell}   \affiliation{STFC Daresbury, Daresbury, Warrington WA4 4AD, UK}		

\author{P.~Rahkila}   \affiliation{University of Jyvaskyla, Department of Physics, P.O. Box 35, FI-40014 University of Jyvaskyla, Finland}		
                      \affiliation{Helsinki Institute of Physics, University of Helsinki, P.O. Box 64, FIN-00014 Helsinki, Finland }

\author{E.~Rapisarda}   
\affiliation{CERN, CH-1211 Geneva 23, Switzerland}	
\affiliation{Paul Scherrer Institut, Villigen, Switzerland}	

\author{A.~\c{S}erban}   \affiliation{``Horia Hulubei" National Institute for Physics and Nuclear Engineering, RO-077125 Bucharest, Romania}

\author{C.O.~Sotty}   
\affiliation{``Horia Hulubei" National Institute for Physics and Nuclear Engineering, RO-077125 Bucharest, Romania}	

\author{L.~Stan}   \affiliation{``Horia Hulubei" National Institute for Physics and Nuclear Engineering, RO-077125 Bucharest, Romania}

\author{M.~St\u{a}noiu}   \affiliation{``Horia Hulubei" National Institute for Physics and Nuclear Engineering, RO-077125 Bucharest, Romania}	

\author{O.~Tengblad} \affiliation{Instituto de Estructura de la Materia, CSIC, Serrano 113 bis, E-28006 Madrid, Spain}

\author{A.~Turturic\u{a}}   \affiliation{``Horia Hulubei" National Institute for Physics and Nuclear Engineering, RO-077125 Bucharest, Romania}

\author{P.~Van~Duppen}   \affiliation{KU Leuven, Instituut voor Kern- en Stralingsfysica, Celestijnenlaan 200D, 3001 Leuven, Belgium}

\author{N.~Warr}   \affiliation{Institut f\"ur Kernphysik, Universit\"at zu K\"oln, Z\"ulpicher Strasse 77, D-50937 K\"oln, Germany}

%% IS530 2012 co-authors

\author{Ph.~Dessagne}
  \affiliation{IPHC, Universit\'e de Strasbourg, IN2P3/CNRS; BP28, F-67037 Strasbourg Cedex, France}

\author{T.~Stora}
  \affiliation{ISOLDE/CERN, Geneva, Switzerland}

\author{C. Borcea}
  \affiliation{``Horia Hulubei" National Institute for Physics and Nuclear Engineering, RO-077125 Bucharest, Romania}

\author{S.~C\u{a}linescu}
  \affiliation{``Horia Hulubei" National Institute for Physics and Nuclear Engineering, RO-077125 Bucharest, Romania}
\author{J.M.~Daugas}
  \affiliation{CEA, DAM, DIF, F-91297 Arpajon Cedex, France}

\author{D.~Filipescu}
  \affiliation{``Horia Hulubei" National Institute for Physics and Nuclear Engineering, RO-077125 Bucharest, Romania}

\author{I.~Kuti}
  \affiliation{Institute for Nuclear Research, H-4001 Debrecen, Pf.51, Hungary}

\author{S.~Franchoo}
  \affiliation{IPNO, Universit\'e Paris-Sud 11, CNRS/IN2P3, Orsay, France}

\author{I.~Gheorghe}
  \affiliation{``Horia Hulubei" National Institute for Physics and Nuclear Engineering, RO-077125 Bucharest, Romania}

\author{P.~Morfouace}
  \affiliation{IPNO, Universit\'e Paris-Sud 11, CNRS/IN2P3, Orsay, France}
  \affiliation{CEA, DAM, DIF, F-91297 Arpajon Cedex, France}

\author{P.~Morel}
  \affiliation{CEA, DAM, DIF, F-91297 Arpajon Cedex, France}

\author{J.~Mrazek}
  \affiliation{Nuclear Physics Institute of the Czech Academy of Sciences, CZ-25068 Rez, Czech Republic}

\author{D.~Pietreanu}
  \affiliation{``Horia Hulubei" National Institute for Physics and Nuclear Engineering, RO-077125 Bucharest, Romania}

\author{D.~Sohler}
  \affiliation{Institute of Nuclear Research, H-4001 Debrecen, Pf.51, Hungary}

\author{I.~Stefan}
  \affiliation{IPNO, Universit\'e Paris-Sud 11, CNRS/IN2P3, Orsay, France}

\author{R. \c{S}uv\u{a}il\u{a}}
  \affiliation{``Horia Hulubei" National Institute for Physics and Nuclear Engineering, RO-077125 Bucharest, Romania}

\author{S. Toma}
  \affiliation{``Horia Hulubei" National Institute for Physics and Nuclear Engineering, RO-077125 Bucharest, Romania}

\author{C.A. Ur}
  \affiliation{``Horia Hulubei" National Institute for Physics and Nuclear Engineering, RO-077125 Bucharest, Romania}
  \affiliation{Istituto Nazionale di Fisica Nucleare, Padova, Italy}

\begin{abstract}

The structure of $^{34}$Si was studied through $\gamma$ spectroscopy separately in the $\beta^-$ decays of $^{34}$Mg and $^{34}$Al at the ISOLDE facility of CERN. 
Different configurations in $^{34}$Si were populated independently from the two recently identified $\beta$-decaying states in $^{34}$Al having spin-parity assignments $J^\pi = 4^-$ dominated by the normal configuration $\pi (d_{5/2})^{-1} \otimes \nu (f_{7/2})$ and $J^\pi = 1^+$ by the intruder configuration $\pi (d_{5/2})^{-1} \otimes \nu (d_{3/2})^{-1}(f_{7/2})^{2}$. The paper reports on spectroscopic properties of $^{34}$Si such as an extended level scheme, spin and parity assignments based on log($ft$) values and $\gamma$-ray branching ratios, absolute $\beta$ feeding intensities and neutron emission probabilities. A total of 11 newly identified levels and 26 transitions were added to the previously known level scheme of $^{34}$Si. 
Large scale shell-model calculations using the {\sc sdpf-u-mix} interaction, able to treat higher order intruder configurations, are compared with the new results and conclusions are drawn concerning the predictive power of {\sc sdpf-u-mix}, the $N=20$ shell gap, the level of mixing between normal and intruder configurations for the 0$_1^+$, 0$_2^+$ and 2$_1^+$ states and the absence of triaxial deformation in $^{34}$Si.

\end{abstract}


\maketitle

\thispagestyle{empty} \pagestyle{empty}

%---------------------------------------------------------------------------------------------------------------------
%RL: since we decided for PRC and now have enough space, I added some of the previous paragraphs in the introduction to give a better overview of the region.

\section{Introduction}
 
Nuclear deformation and shape coexistence have been a topic of interest in nuclear structure research for more than five decades \cite{heyde}. An increasing flow of experimental data proved the robustness of the magic shell closures around the stability line, and revealed their weakening or the development of new ones while going far from stability. A particular manifestation of such phenomena in the exotic nuclei landscape is the occurrence of 'Islands of Inversion' \cite{Pove87, Warb90,Heyd91,Fuku92, PPNP}. It was shown that for nuclei with $N$= 20, 28, 40 (and their vicinity) while changing the proton number $Z$ for a given neutron number $N$, nuclear structure properties no longer agree with the 'closed neutron shell' predictions.

The experimental findings were gradually understood and theoretically explained: the balance between shell and sub-shell energy gaps (an independent-particle effect) and large correlation energy (mainly due to pairing and quadrupole two-body forces) are key for the understanding of shape coexistence in nuclei.  In a shell-model framework these phenomena can be seen as a consequence of the presence of multiparticle - multihole ($np-nh$) configurations in the ground states of nuclei such as $^{32}$Mg, $^{42}$Si, $^{64}$Cr \cite{caurier2, sorlin, Pov2017}.  For particular $Z$ and $N$ values, the correlation energy in these intruder configurations (quadrupole and pairing energy) is higher than in the case of normal configurations (no particle-hole excitations). This effect, combined with a lowered shell-gap, leads to ground-states in both even-even and odd-mass nuclei that have strongly correlated states \cite{smirnova}. 

The previously known level scheme of $^{34}$Si comprising of 8 excited states and 10 transitions was established through experiments at the CERN online mass separator ISOLDE by studying the $\beta$-decays of $^{34,35}$Al \cite{nummela, baumman}, at the NSCL facility of MSU through the $^7$Li$(^{34}$P,$^7$Be$+ \gamma$) reaction \cite{zegers}, and other various experiments \cite{Ende02, For94, fifield, Iwa03, ibbotson}. Due to its closed-shell $Z$= 14, $N$= 20 character, $^{34}$Si has the properties of a doubly-magic spherical nucleus (e.g. high 2$^+$ energy \cite{baumman}, low $B(E2; 2_1^+ \rightarrow  0_{1}^+)$ value \cite{ibbotson}, drop in $S_n$ value after $N=20$), but lies at the verge of the 'Island of Inversion', where nuclei are deformed in their ground state configuration. It follows that deformed configurations, shape coexistence \cite{rotaru} and possibly triaxial shapes are present already among the few first excited states in $^{34}$Si \cite{Han2017}. The abrupt transition from the closed-shell ground-state of $^{34}$Si to the intruder-dominated deformed ground state of $^{32}$Mg \cite{wimmer}, while removing only two protons in the $1d_{5/2}$ orbit, is a challenge for nuclear models \cite{Uts2012, Bor2015}. This is due to the delicate balance between the amplitude of the proton and neutron shell gaps that prevents nuclear excitations, and the large correlation energy that are maximum when many particle-hole excitations across these gaps are present. A central proton density depletion, so-called 'bubble', \cite{Muts16} was recently identified in $^{34}$Si, one of the few nuclei that experience a drastic reduction of its spin-orbit interaction (for $L=1$ neutrons), as compared to the neighboring nuclei.

$^{34}$Si was recently studied at GANIL \cite{rotaru}  through the $\beta$-decay of $^{34}$Al produced in the 'one neutron pick-up and three proton removal' reaction channel using a $^{36}$S beam at intermediate energy (77.5\,MeV/A). Through electron-positron pairs energy measurements, the authors identified the 0$_2^+$ state in $^{34}$Si at 2.7 MeV excitation energy, with a half-life of T$_{1/2}=$\,19.3(7)\,ns and also determined the reduced monopole transition strength $\rho^2(E0, 0_2^+ \rightarrow 0_{1}^+) = 13(1)\times10^{-3}$.  It was the first observation of the $0_2^+$ state which showed a $2\hbar\omega$ intruder character with predicted oblate deformation, while the ground state was spherical (closed shell configuration). The measured $\rho^2(E0)$ allowed probing the shape coexistence. Among other observables, the branching ratio in the de-excitation of the $2_1^+$ state of $^{34}$Si and the $B(E2; 2_1^+ \rightarrow 0_2^+)$ is directly related to the degree of deformation and shape mixing.  In the aforementioned GANIL experiment, this branching was deduced with a very large relative uncertainty of 50\% due to the statistical fluctuations in the background subtraction and in the detection efficiencies for electron-positron pairs emitted with an unknown energy-angle correlation. 

The present paper addresses some aspects of the $^{34}$Si nucleus structure that remain poorly known, still questionable, and that are essential benchmarks for theoretical calculations on doubly-magic nuclei in general: the amount of mixing between the first two 0$^+$ states (that we shall determine more accurately as compared to Ref.\,\cite{rotaru}),  the size of the $N=20$ gap determined through the study of neutron excitations, the identification of the spherical  2$^+$ state (as a confirmation of Ref. \cite{zegers}), and the possible existence of trixiality, as predicted by the Gogny D1S and SDPF-M interactions and suggested experimentally in Ref.~\cite{Han2017}.

To achieve these challenging goals, we have studied the combined $\beta$-decay of the $^{34}$Mg and $^{34}$Al nuclei, that has the advantage to provide information on positive and negative parity states in $^{34}$Si over a broad range of energy up to the neutron emission threshold, $S_n = 7.514(15)$~MeV. A total of 11 newly identified levels with tentative spin and parity assignments and 26 transitions were added to the previously known level scheme of $^{34}$Si.

%---------------------------------------------------------------------------------------------------------------------

\section{Experiment} 

The $^{34}$Mg and $^{34}$Al ions were produced at the ISOLDE-CERN facility through fragmentation reactions induced by a 1.4\,GeV pulsed proton beam delivered by the PS-Booster, with an average intensity of 1.9\,$\mu$A, on a standard UC$_x$ target. After being accelerated by a 40\,kV potential and selectively ionized using the Resonance Ionization Laser Ion Source technique (RILIS) \cite{rilis, koester2}, the $A=34$  Mg or Al elements were selected by the ISOLDE General Purpose Separator (GPS) and implanted on a movable tape, located at the center of the ISOLDE Decay Station (IDS) \cite{ids}.

\begin{figure}[!h]
\begin{minipage}[t]{\linewidth}
    \centering

\includegraphics[width=8.6 cm]{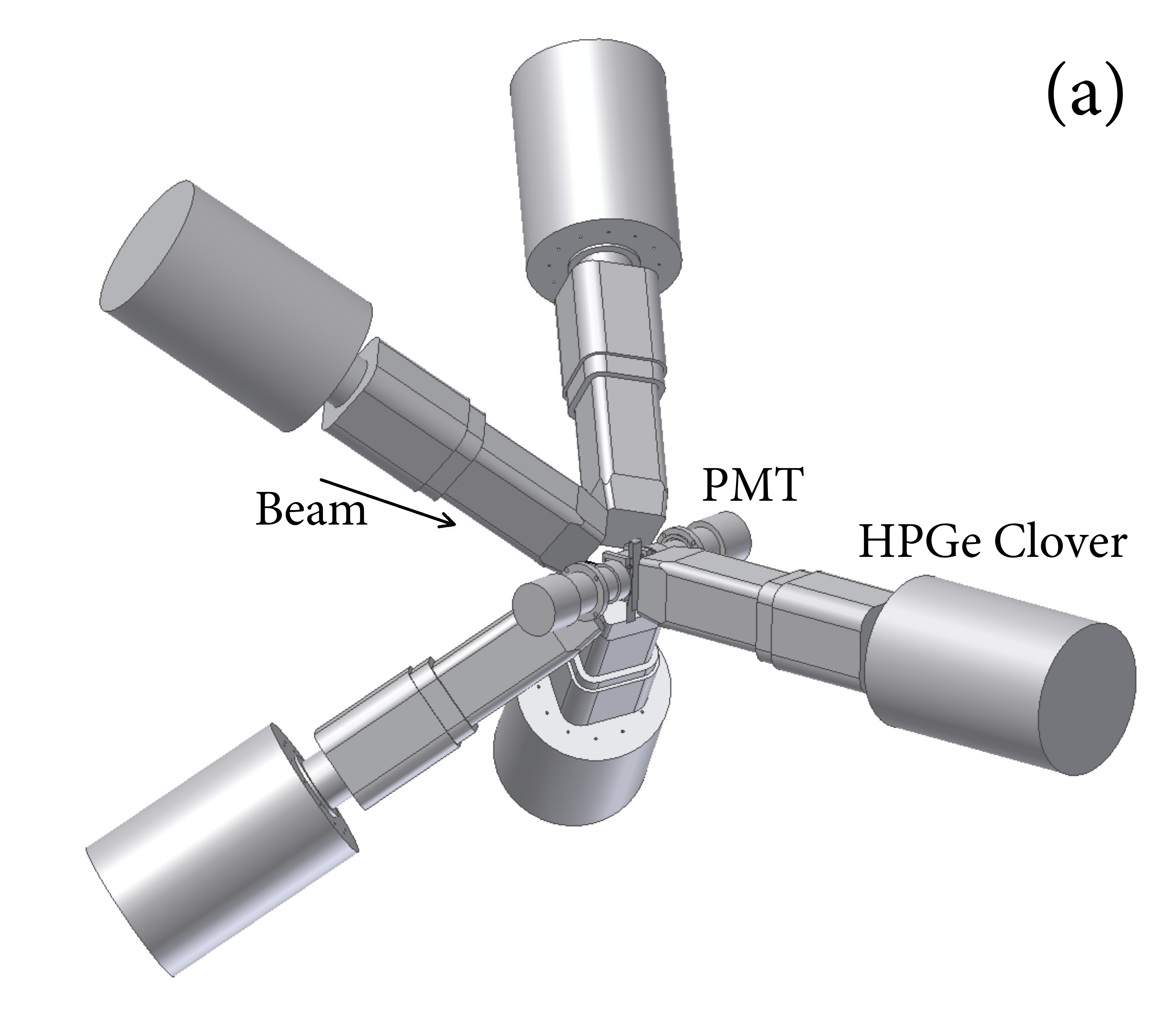}
\includegraphics[width=5.8 cm]{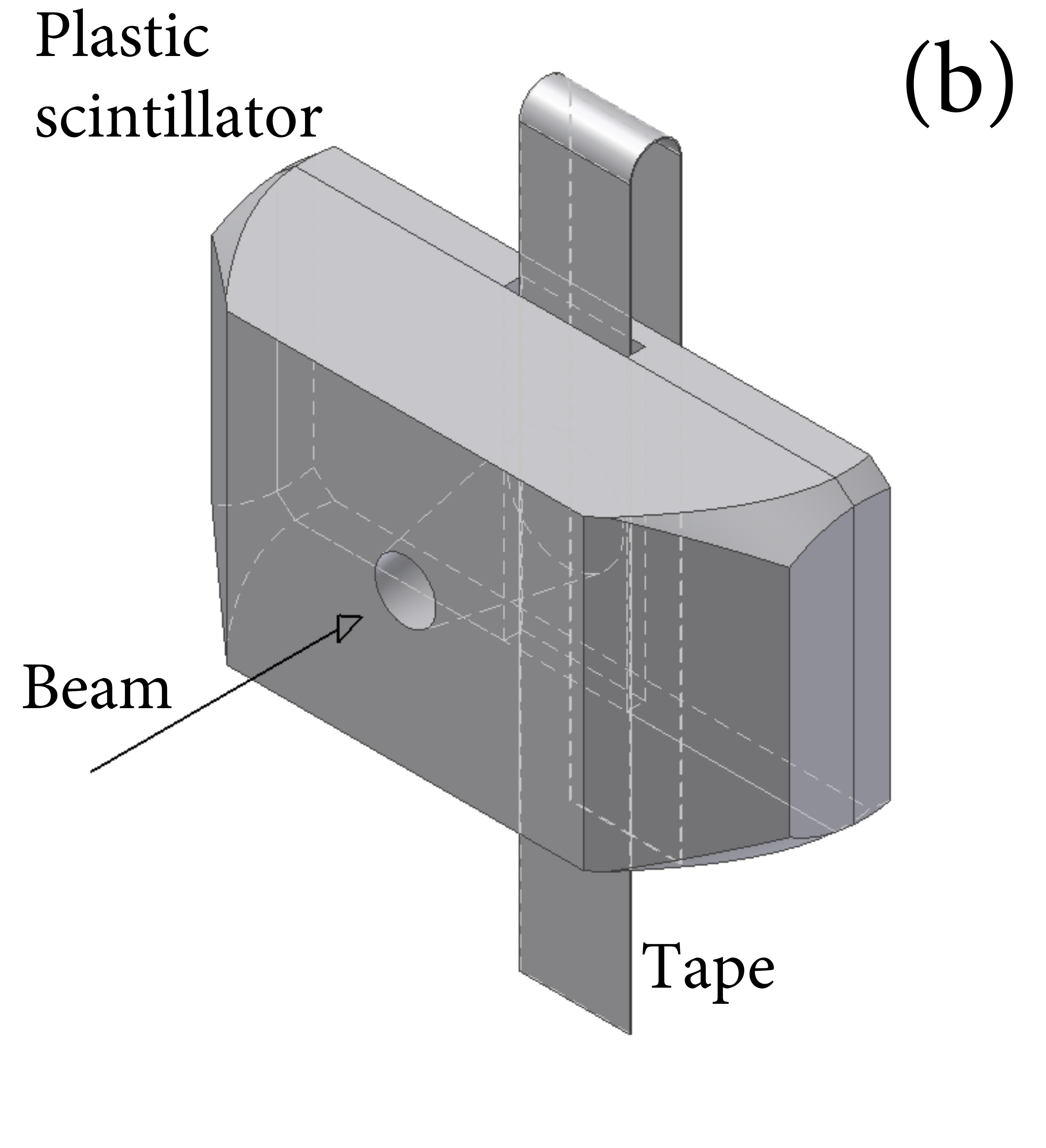}

\caption{\label{fig:setup} Graphical representation of the high $\beta$-$\gamma$ efficiency configuration of IDS. (a) Five HPGe Clover detectors were placed in close geometry, four at $\sim$75\,mm and one at $\sim$60\,mm from the implantation point and two PMTs were used to read out the plastic scintillator. (b) Detailed view of the plastic scintillator. The beam passed through a 10\,mm opening in the plastic scintillator and was implanted in the aluminized mylar tape which was moved periodically in order to remove the long-lived daughter activity.}
\end{minipage}        
\end{figure}

Quasi-pure beams of $^{34}$Mg and $^{34}$Al (with purities $>$ 99\%) were obtained with intensities of $7(1)\times10^2$ and $8(1)\times10^2$ \,ions/s, respectively, leading to a total of $\sim7\times10^7$ implanted ions of either beam over the whole experiment. Two other settings were used to select the $^{33}$Mg and $^{33}$Al nuclei in order to determine absolute decay intensities in the $^{34}$Mg and $^{34}$Al decay chains.

The ISOLDE beam gate was open during an adjustable time gate after every proton pulse (which occurred at intervals of 1.2 s), during which the nuclei of interest were continuously implanted on the movable tape. The $\beta$-decay measurements took place at the collection point during the implantation and the subsequent decay. To remove the long-lived daughter activity, the tape was moved at a certain time interval after each proton pulse. Data were collected for each radioactive ion beam with optimized time gates (50 -- 200\,ms) and tape transport conditions (around 500 ms after the proton pulse or no transport at all) to determine level schemes, absolute intensities and $\beta$-decay half-lives for the isotopes of interest. 

The detection setup, shown in Fig.\,\ref{fig:setup}(a), is the same as the one briefly described in Ref.\,\cite{Lica2017}. It represents the high $\beta$-$\gamma$ efficiency configuration of IDS.  
$\gamma$ rays were detected in five HPGe Clover detectors arranged in a close geometry at $\sim$7~cm from the implantation point, leading to efficiencies of 6\% at 600\,keV and 3\% at 2000\,keV, after the add-back procedure \cite{duchene} was employed. The $\gamma$-ray photopeak efficiency of the HPGe detectors was determined using the $^{152}$Eu calibration source and extrapolated using GEANT4 \cite{geant4} simulations. 
$\beta$ particles were detected in a 3\,cm-thick NE102 plastic scintillator, shown in Fig.\,\ref{fig:setup}(b), which was made out of two joined pieces that covered a solid angle of $\sim$95\% around the implantation point. Signals induced in the plastic scintillator were read simultaneously by 2 photomultiplier tubes (PMTs) placed at opposite ends. Only the events that triggered both PMTs were considered, which allowed the energy thresholds to be lowered near the level of the phototube dark current in order to reach a $\beta$ efficiency close to the geometrical value. 
The 90(5)\% $\beta$ efficiency of the plastic scintillator was determined from the ratio between various $\beta$-gated and singles $\gamma$ rays, in agreement with the ratio between the total number of $\beta$ decays recorded and expected in the full decay chains of $^{33,34}$Mg and $^{33,34}$Al when the tape was not moved after source collection. 
All the signals were recorded and sampled in a self-triggered mode using the 14-bit 100\,MHz Nutaq VHS-V4 data acquisition (DAQ) system of the IDS \cite{nutaq}. The digital processing of the energy signals provides resolutions  at E$_\gamma$=1.3~MeV of the order of 2.3~keV for the HPGe detectors. A 10-bit 1\,GS/s V1751 Caen  digitizer was used to record signal traces from the plastic scintillator, offering the possibility to detect $\beta$ -- ($e^- e^+$) coincidences down to $\simeq$\,10\,ns time range. Both systems were synchronized in order to recover coincidences offline.

%(four at 75\,mm and one at 60\,mm from the implantation point) 
% 
%
%                      26mm    |             Beam Direction
%   implantation   o-----------|10mm            <------ 
%      point                   |(openning)
%
% solid angle = 2*pi*(1-26/sqrt(26^2+5^2)) = 2*pi*(1-0.982)
% the upper and lower openning through which the tape is passing amount for another ~2%
%
%

%---------------------------------------------------------------------------------------------------------------------

\subsection{Level Scheme of $^{34}$Si}

The decays of $^{34}$Mg and $^{34}$Al were used to populate low-spin positive-parity and high-spin negative-parity states in $^{34}$Si, respectively. Indeed, it was observed in Ref.~\cite{Lica2017} that more than 99\% of the $\beta$ decay of $^{34}$Mg ($T_{1/2}=44.9(4)$\,ms) proceeds through the $1^+$ isomeric state of $^{34}$Al, which subsequently populates mainly the low-spin positive parity states in $^{34}$Si. As for the $^{34}$Al beam, 89(3)\% of ground (4$^-$,  $T_{1/2}=53.73(13)$\,ms) and 11(3)\% of isomeric (1$^+$, $T_{1/2}=22.1(2)$\,ms) state were deduced from transitions in $^{34}$Si decaying from high-spin states (e.g. $(5^-) \rightarrow (4^-)$ 590.8 keV) and low-spin states (e.g. $0_2^+ \rightarrow 0_1^+$ 2718.4 keV E0), respectively. The  ratio between the ground and isomeric state population in the $^{34}$Al beam is determined by the reaction mechanism, the target characteristics and the ion-source employed.

\begin{figure*}[htbp]
\includegraphics[width=14cm]{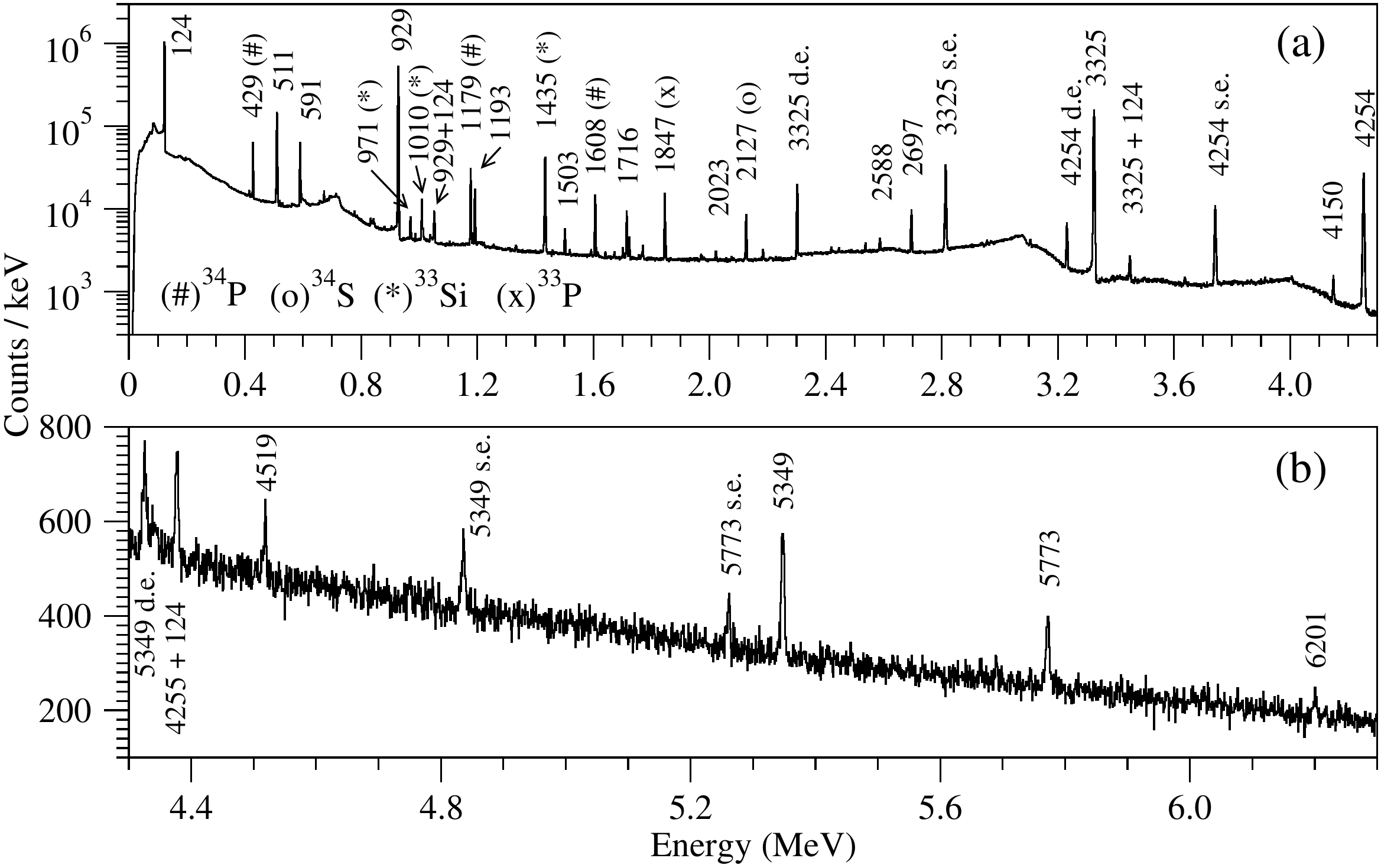}
\caption{\label{fig:E-Clov-Beta}  $\beta$-gated $\gamma$-ray HPGe spectrum of the $^{34}$Al decay recorded during the first 500\,ms after the proton impact. The energy ranges shown are (a) 0 - 4.3 MeV and (b) 4.3 - 6.3 MeV. The most intense transitions in $^{34}$Si are labeled together with the ones originating from longer lived daughter nuclei. The latter are significantly suppressed because of the short 500\,ms gating requirement and subsequent movement of the tape. The peaks corresponding to the $\gamma$ rays from the long-lived daughters are indicated with symbols: (\#) $^{34}$P, (o) $^{34}$S, ($\ast$) $^{33}$Si, (x) $^{33}$P. No contaminants could be identified. }
\end{figure*}

\begin{figure}[!h]
\begin{minipage}[t]{\linewidth}
    \centering

    \includegraphics[width=8.6 cm]{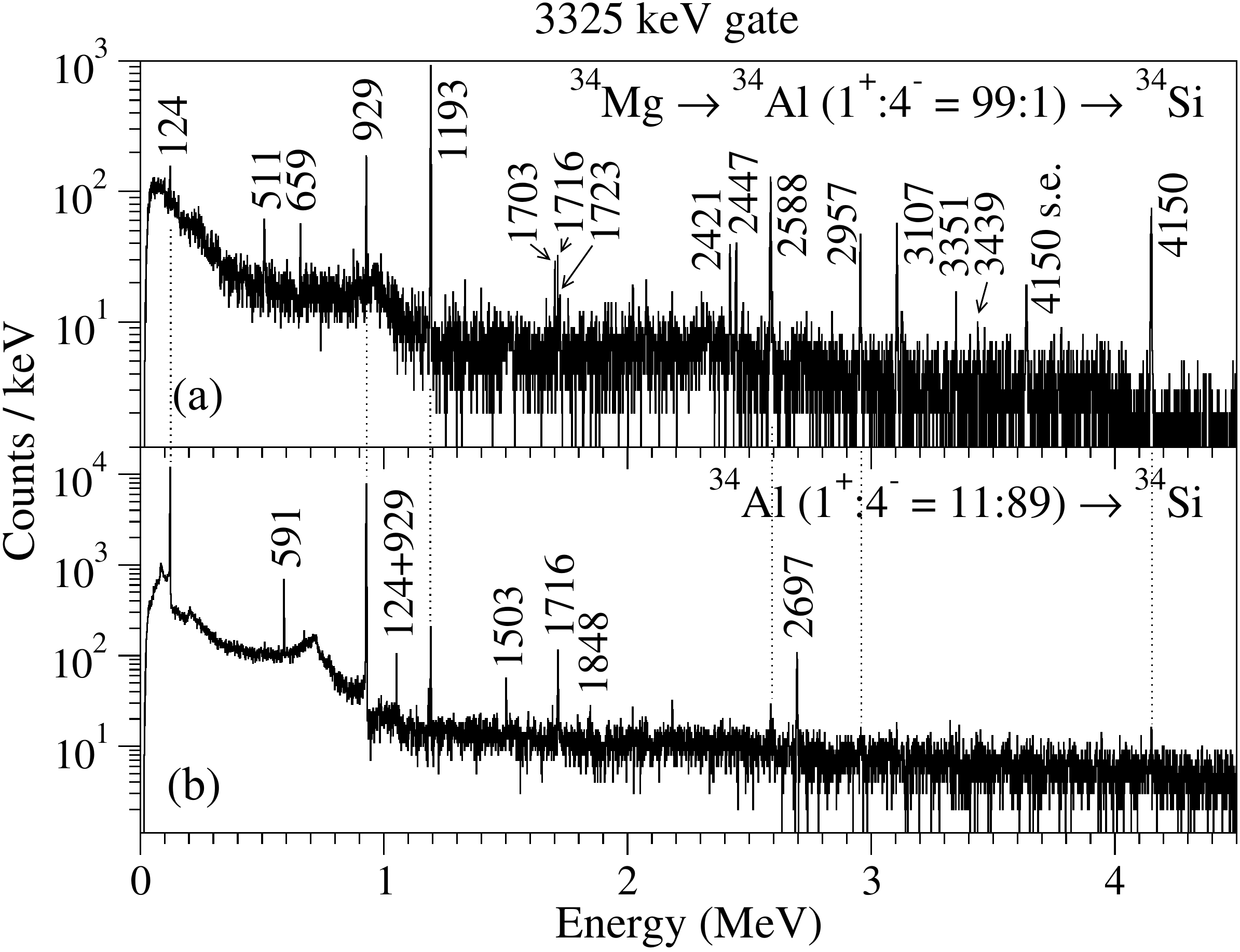}
   
\caption{\label{fig:3325gate} Background-subtracted, $\beta$-gated $\gamma$-ray spectra in coincidence with 3325-keV $\gamma$ rays obtained from the decay of implanted  (a) $^{34}$Mg and (b) $^{34}$Al  isotopes. The different ratios in which the 4$^-$ and 1$^+$ states in $^{34}$Si were populated are indicated in brackets and explained in the text.}
\end{minipage}        
\end{figure}

\begin{figure*}[htbp]
\includegraphics[angle=-90, width=15.5cm]{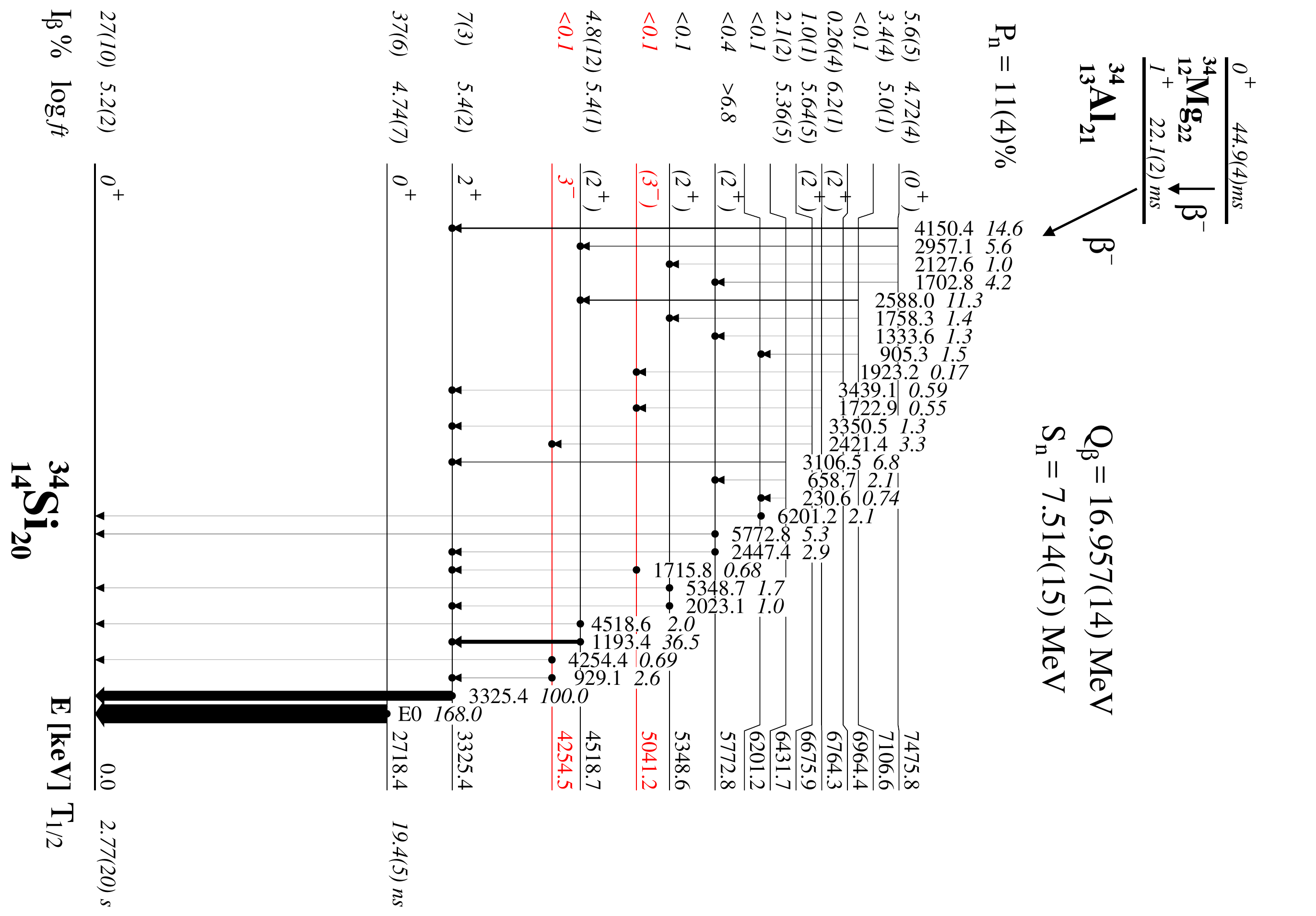}
\includegraphics[angle=-90, width=15.5cm]{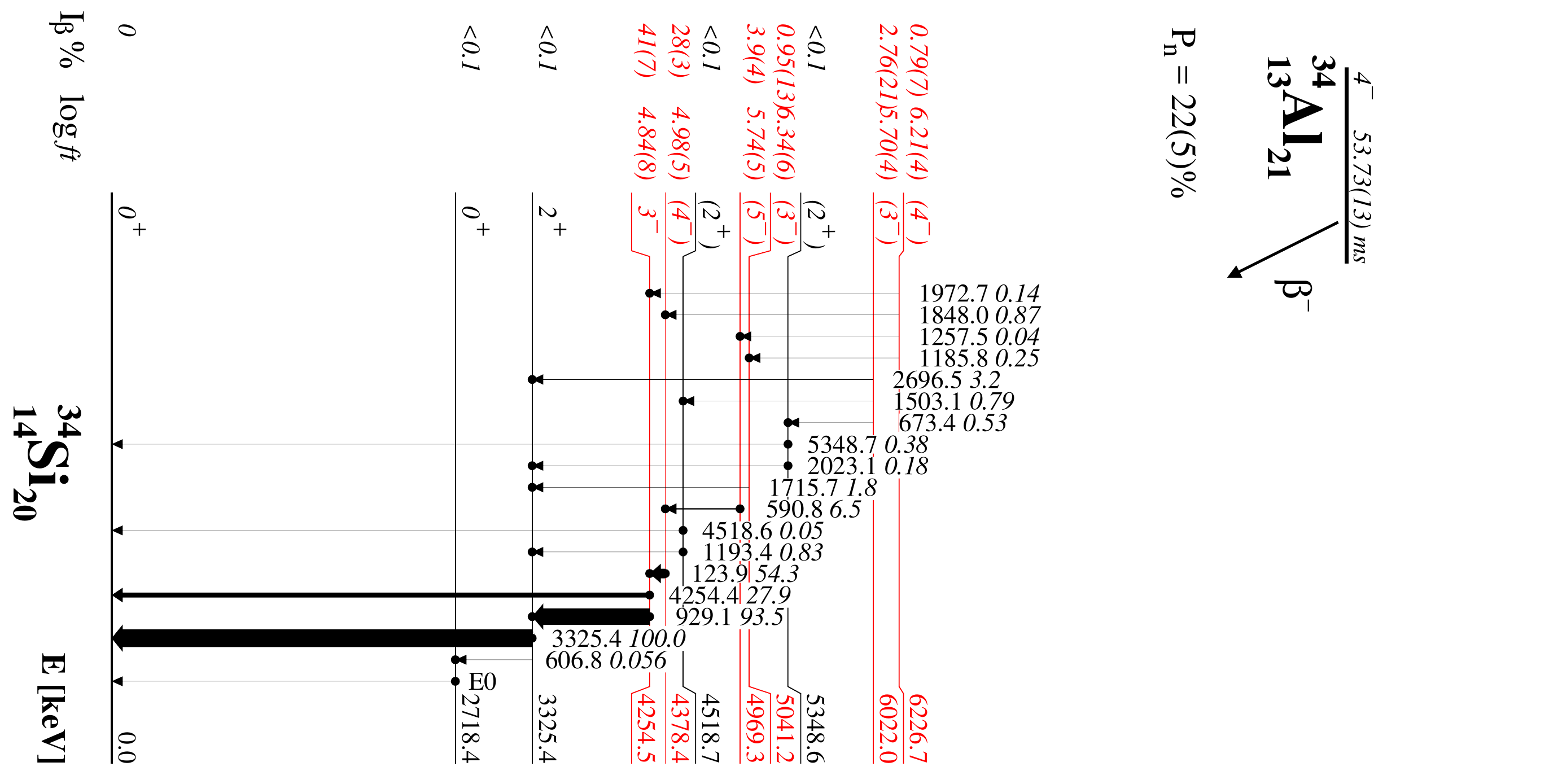}
\includegraphics[angle=-90, width=15.5cm]{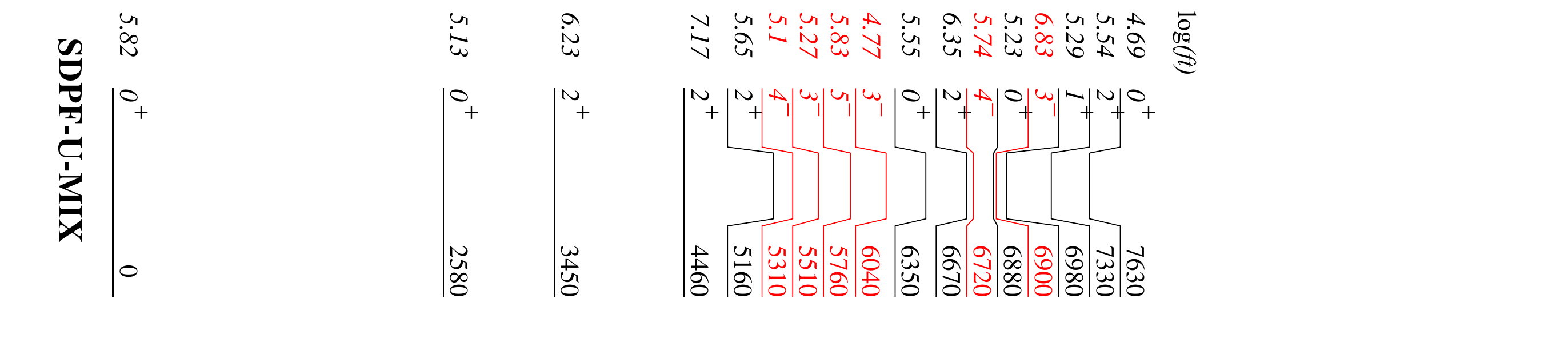}
\caption{\label{fig:levelScheme} (Color Online) Level schemes of $^{34}$Si populated separately in the $\beta$ decay of the $4^-$ ground state and $1^+$ isomer of $^{34}$Al and shell-model calculation using the {\sc sdpf-u-mix} interaction \cite{caurier}. A total of 11 newly identified levels and 26 transitions were added to the previously known level scheme of $^{34}$Si. The levels indicated in red are assumed to have negative parity. Tentative spins and parities are assigned based on $\gamma$-ray branching ratios, log($ft$) values and comparisons to shell-model calculations. The $\gamma$-ray intensities are relative to the 3325.4-keV transition. For absolute intensity per 100 decays, multiply by 0.22(3) and 0.61(6) for the decay of the $1^+$ and $4^-$ states in $^{34}$Al, respectively.  }
\end{figure*}

The $\beta$-gated $\gamma$-ray energy spectrum of $^{34}$Al, recorded in the first 500\,ms from the proton beam impact on the ISOLDE target, is shown Fig.\,\ref{fig:E-Clov-Beta}. 
%Data was collected continuously, however the tape was moved 500\,ms after every proton impact in order to remove the daughter activity. 
$\beta$-$\gamma$ and $\beta$-$\gamma$-$\gamma$ coincidences for the decay of $^{34}$Mg and $^{34}$Al, some of which are shown in Fig.\,\ref{fig:3325gate}, are used to establish the level scheme of $^{34}$Si shown in the left and right part of Fig.\,\ref{fig:levelScheme}, respectively. When $\beta$-$\gamma$-$\gamma$ coincidences could not be used, the placement of transitions in the level scheme is based on their relative intensity and energy matching conditions. The determination of the excitation energy of the levels obtained from different $\gamma$-ray cascades agrees within 0.3\,keV.  No recoil correction has been applied to the $\gamma$-ray energies, since the studied isotopes were implanted into the tape and recoil effects were estimated to be lower than the corresponding $\gamma$-ray energy measurement error ($<$0.3\,keV). 

The assignment of tentative spins and parities in the level scheme is based on log($ft$) values for allowed Gamow-Teller (GT) transitions, as well as from $\gamma$-ray branching ratios decaying from or to levels with known spins and parities. Comparison to shell-model calculations with the {\sc sdpf-u-mix} interaction will be used as an additional guidance. The log($ft$) values are derived from partial decay lifetimes which make use of the level $\beta$ feeding (for which the number of implanted nuclei is needed) and the total $\beta$-decay lifetime. The ground-state-to-ground-state $^{34}$Al $Q_{\beta}$ value of 16.957(14)\,MeV \cite{ame2016} was used when deriving the log($ft$) values, considering also the 46.7 keV excitation energy of the $\beta$ decaying isomeric state in $^{34}$Al \cite{Lica2017}.
%The main  observed transitions  at 123.9 and 929.1 keV were previously reported in \cite{nummela} as being populated in the $\beta$-decay of the $^{34}$Al $4^-$ ground state.

The number of implanted $^{34}$Mg and $^{34}$Al nuclei is derived from singles $\gamma$-ray spectra, using the transitions corresponding to the last populated daughters in the decay chains. The beam-gate was reduced to 100\,ms and the tape was not moved in order to study the full decay chains of separately implanted $^{33,34}$Mg and $^{33,34}$Al. Consistent absolute intensities were extracted successively by analysing the full decay chains, using only the following literature values for absolute $\gamma$-ray intensities and neutron emission probabilities: $P_n$($^{33}$Mg) $=14(2)\%$ \cite{angelique}, $P_n$($^{33}$Al) $=8.5(7)\%$ \cite{reeder}, $I_{abs}$(1618\,keV; $^{33}$Mg $\rightarrow$ $^{33}$Al) $=16(2)\%$ \cite{tripathi}. An upper limit for the two-neutron emission probability of $^{34}$Mg, $P_{2n}<0.1\%$, was determined. The main absolute $\beta$ and $\gamma$ intensities extracted for $^{34}$Si levels populated from  the $\beta$ decay of the $1^+$ or $4^-$ states are given in Tables\,\ref{tab:1+} and \ref{tab:4-}. In Table\,\ref{Tab1} are indicated absolute intensities for the most intense transitions in isotopes from the  $^{34}$Mg decay chain, as determined in the present experiment.

%from the deduced intensities of the $\gamma$ rays populating excited states in daughter nuclei ($I_{abs}($1618\,keV; $^{34}$Mg $\rightarrow$ $^{33}$Al$)=13(5)\%$, $I_{abs}($364\,keV; $^{34}$Mg $\rightarrow$ $^{34}$Al)$=11.2(15)\%$).

\begin{figure}[!h]
\begin{minipage}[t]{\linewidth}
    \centering

    \includegraphics[width=8.6 cm]{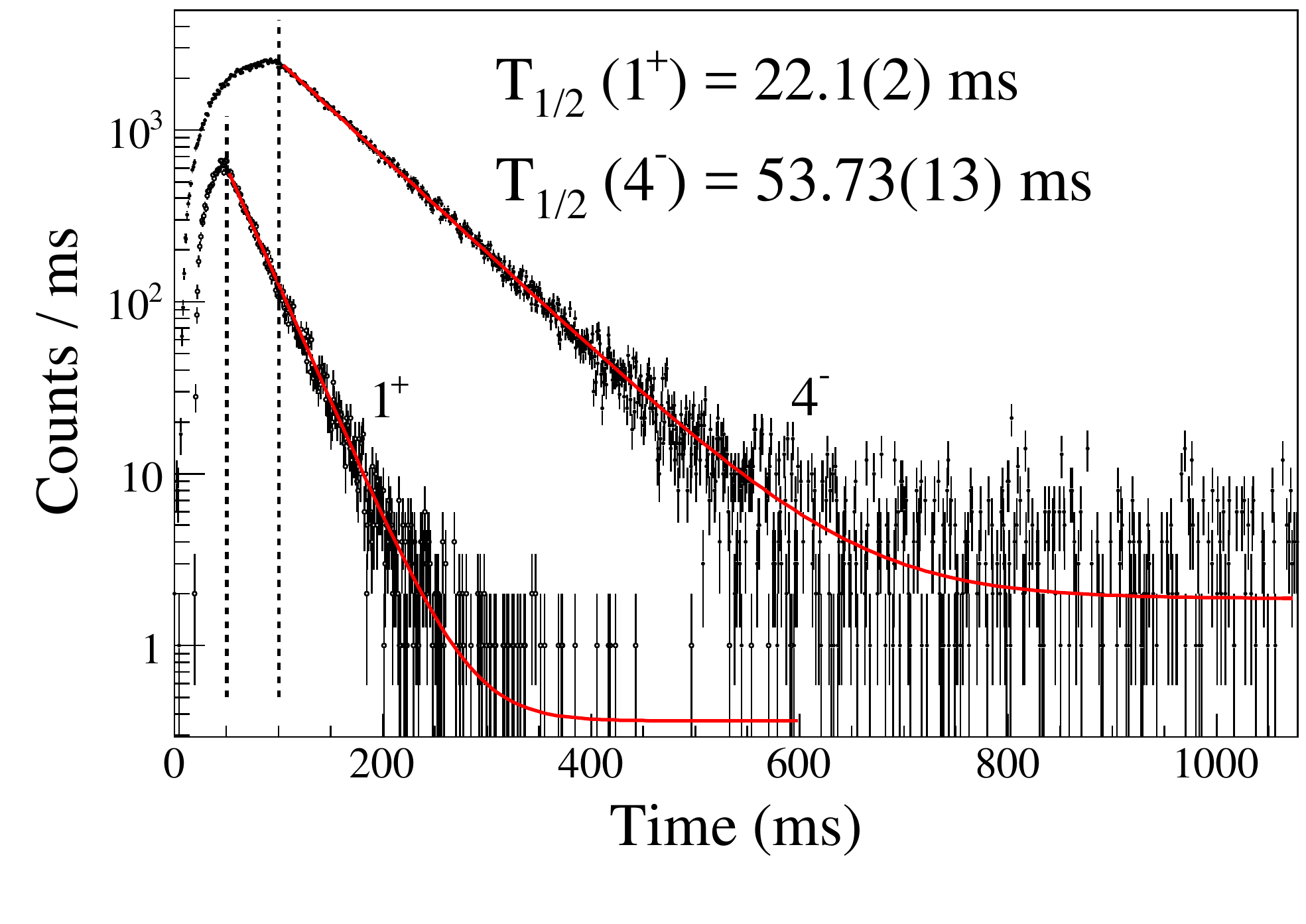}
   
\caption{\label{fig:halflives} (Color Online)  The half-lives of the decaying 1$^+$ and 4$^-$ states, $T_{1/2} = 22.1(2)$\,ms and $53.73(13)$\,ms, respectively, were extracted from the time distributions of transitions in $^{34}$Si: $E0(0_2^+\rightarrow0_1^+)$ and $E1(929.1$\,keV; $3^-\rightarrow2^+)$, respectively. 
 }
\end{minipage}        
\end{figure}

\begin{figure}[!h]
\begin{minipage}[t]{\linewidth}
    \centering

    \includegraphics[width=8.6 cm]{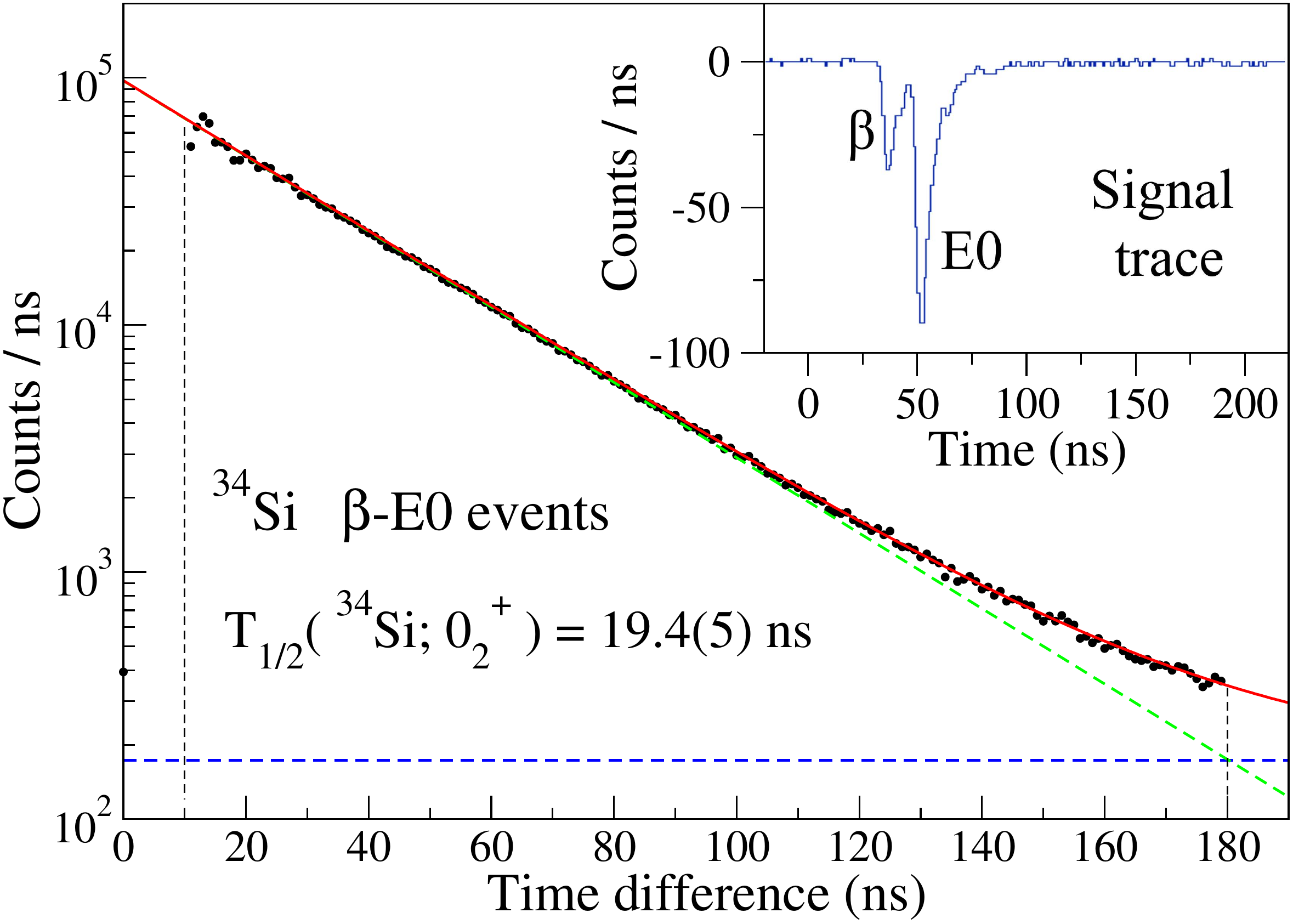}
   
\caption{\label{fig:34Mg_fit} (Color Online) Time difference distribution of double events ($\beta$ - E0) in the plastic scintillator. The inset shows a signal trace of a double event: the $\beta$ electron followed closely by the $e^-e^+$ signal originating from the $0_2^+ \rightarrow 0_1^+$ E0 transition in $^{34}$Si. The half-life $T_{1/2} = 19.4(5)$\,ms of the $0_2^+$ state in $^{34}$Si was extracted after fitting the distribution using a sum between an exponential function and a constant representing the background.  }
\end{minipage}        
\end{figure}

An accurate determination of the $\beta$-decay half-life of the 1$^+$ isomer and 4$^-$ ground state of $^{34}$Al was obtained by reducing the beam gate to $50$\,ms and $100$\,ms, respectively. 
As first reported in \cite{rotaru}, the 2718-keV $0_2^+$ state in $^{34}$Si decays mainly through internal pair formation (IPF) towards the $0_1^+$ ground state. The internal conversion (IC) transition rate is negligible, having a contribution of only 0.5~\% to the total rate.  The $e^-e^+$ pairs originating from the 2718-keV  $E0(0_2^+\rightarrow0_1^+)$ transition were detected in delayed coincidence with $\beta$ particles, as double events in the plastic scintillator (shown in Fig.\,\ref{fig:34Mg_fit}). 
The half-lives of the decaying 1$^+$ isomer ($22.1(2)$\,ms) and 4$^-$ ground state ($53.73(13)$\,ms), were extracted from the time distributions of transitions in $^{34}$Si: $E0(0_2^+\rightarrow0_1^+)$ and $E1(929.1$\,keV; $3^-\rightarrow2^+)$, respectively, as shown in Fig.\,\ref{fig:halflives}. The present values are more precise, but consistent with the values obtained in Refs. \cite{rotaru,Han2017}.

%An important result concerning the level scheme of $^{34}$Si is the first-time identification of its excited states populated separately in the decay of the 4$^-$ and 1$^+$ $\beta$-decaying states of $^{34}$Al. This allowed for the clear identification of high-spin negative parity and low-spin positive parity states in $^{34}$Si, based on the measured $\gamma$-decay branching ratios and comparisons with the calculated level scheme using the {\sc sdpf-u-mix} interaction. 

\begin{table}[!h]
\caption{ \label{tab:1+} Levels of $^{34}$Si populated in the $\beta$ decay of the $1^+$ isomer of $^{34}$Al. The first columns from the left reports level energies in keV and proposed spins and parities. Columns 3 and 4 show the I$_{\beta}$ calculated as discussed in the text and corresponding log($ft$) values. In the last columns we report the energies of the $\gamma$ transitions de-exciting the level, together with their relative intensity and the level to which it decays. The $\gamma$-ray intensities are relative to the 3325.4-keV transition. For absolute intensity per 100 decays, multiply by 0.22(3). }
\begin{tabular}{lrrrrrr}
\tabularnewline

\hline
\hline
\rule{0pt}{3ex}
$E_{i}$  & $J^{\pi}$  & $I_{\beta}$  & log($ft$) & $E_{\gamma}$  & $I_{\gamma}$  & $E_{f}$ \\

[keV]    &            & [\%]         &           &  [keV]        & [\%]          & [keV] \\
\hline
\rule{0pt}{3ex}

%EDIT USING nedit
0.0		& 0$^+$ 	& 27(10)		& 5.2(2)	&    		 &		&		\\
2718.4(1) 	& 0$^+$		& 37(6)		& 4.74(7) 	& 2718.4(E0)	 &	168(3) 	& 0.0		\\
3325.4(1) 	& 2$^+$  	& 7(3)		& 5.4(2)	& 3325.4(1)	 &	100(3) 	& 0.0		\\
4254.5(1) 	& 3$^-$  	& $<$0.1	& 		& 929.1(1) 	 &	2.6(1)  & 3325.4	\\
	 	& 	  	& 		& 		& 4254.4(1)	 &	0.7(1)  & 0.0		\\
4518.7(1) 	& (2$^+$)  	& 4.8(12)		& 5.4(1)	& 1193.4(1)	 &	36.5(8) & 3325.4	\\
	 	& 	  	& 		& 		& 4518.6(1)	 &	2.0(1)  & 0.0  	 	\\
5041.2(2) 	& (3$^-$)  	& $<$0.1	& 		& 1715.8(1)	 &	0.7(1)	& 3325.4	\\
5348.6(2) 	& (2$^+$)  	& $<$0.1	& 		& 2023.1(2)	 &	1.0(1)  & 3325.4        \\
	 	& 	  	& 		& 		& 5348.7(2)	 &	1.6(1)  & 0.0	        \\
5772.8(2) 	& (2$^+$)  	& $<$0.4	& $>$6.8		& 2447.4(2)	 &	2.9(2)  & 3325.4	\\
	 	& 	  	& 		& 		& 5772.8(2)	 &	5.3(2)  & 0.0		\\
6201.2(3) 	& 	  	& $<$0.1	& 		& 6201.2(3)	 &	2.1(1)  & 0.0		\\
6431.7(3) 	&   	& 2.1(2)	& 5.36(5)	& 230.6(1) 	 &	0.7(1)  & 6201.2	\\
	 	& 	  	& 		& 		& 658.7(1) 	 &	2.1(1)  & 5772.8	\\
	 	& 	  	& 		& 		& 3106.5(1)	 &	6.8(2)  & 3325.4	\\
6675.9(3) 	& (2$^+$)  	& 1.0(1)	& 5.64(5)	& 2421.4(1)	 &	3.3(1)  & 4254.5	\\
	 	& 	  	& 		& 		& 3350.5(2)	 &	1.3(1)  & 3325.4	\\
6764.3(3) 	& (2$^+$)  	& 0.26(4)	& 6.2(1)	& 1722.9(2)	 &	0.6(1)  & 5041.2	\\
	 	& 	  	& 		& 		& 3439.1(2)	 &	0.6(1)  & 3325.4	\\
6964.4(3) 	& 	  	& $<$0.1	& 		& 1923.2(2)	 &	0.2(1)  & 5041.2	\\
7106.6(3) 	& 	  	& 3.4(4)	& 5.0(1)	& 905.3(1) 	 &	1.5(1)  & 6201.2	\\
	 	& 	  	& 		& 		& 1333.6(1)	 &	1.3(1)  & 5772.8	\\
	 	& 	  	& 		& 		& 1758.3(1)	 &	1.4(2)  & 5348.6	\\
	 	& 	  	& 		& 		& 2588.0(1)  	 &	11.3(4) & 4518.7	\\
7475.8(3) 	& (0$^+$)  	& 5.6(5)	& 4.72(4)	& 1702.8(1)	 &	4.2(2)  & 5772.8	\\
	 	& 	  	& 		& 		& 2127.6(1)	 &	1.0(1)  & 5348.6	\\
	 	& 	  	& 		& 		& 2957.1(1)	 &	5.6(1)  & 4518.7	\\
	 	& 	  	& 		& 		& 4150.4(1)	 &	14.6(5) & 3325.4	\\

\hline
\hline
\end{tabular}
\end{table}

\begin{table}[!h]
\caption{ \label{tab:4-} Levels of $^{34}$Si populated in the $\beta$ decay of the 
$4^-$ ground state of $^{34}$Al. The $\gamma$-ray intensities are relative to the 3325.4-keV transition. For absolute intensity per 100 decays, multiply by 0.61(6). \\}
\begin{tabular}{lrrrrrr}
\tabularnewline

\hline
\hline
\rule{0pt}{3ex}
$E_{i}$  & $J^{\pi}$  & $I_{\beta}$  & log($ft$) & $E_{\gamma}$  & $I_{\gamma}$  & $E_{f}$ \\

[keV]    &            & [\%]         &           &  [keV]        & [\%]          & [keV] \\
\hline
\rule{0pt}{3ex}

%EDIT USING nedit
0.0		& 0$^+$ 	& 0		& 		&    		 &		&		\\
2718.4(1) 	& 0$^+$		& $<$0.1	& 	 	& E0		 &	$<$0.5 	& 0.0		\\
3325.4(1) 	& 2$^+$  	& $<$0.1		&  	& 606.8(1)	 &    0.056(6) 	& 2718.4	\\
	 	& 	  	& 		& 		& 3325.4(1)	 &	100(2)  & 0.0		\\
4254.5(1) 	& 3$^-$  	& 41(7)		& 4.84(8)	& 929.1(1) 	 &	94(2)   & 3325.4        \\
	 	& 	  	& 		& 		& 4254.4(1)	 &	28(1)   & 0.0	        \\
4378.4(1) 	& (4$^-$)  	& 28(3)		& 4.98(5)	& 123.9(1) 	 &	54(2)   & 4254.5        \\
4518.7(1) 	& (2$^+$)  	& $<$0.1	& 		& 1193.4(1)	 &	0.83(3) & 3325.4	\\
	 	& 	  	& 		& 		& 4518.6(1)	 &	0.05(2) & 0.0  	 	\\
4969.3(1) 	& (5$^-$)  	& 3.9(4)	& 5.74(5)	& 590.8(1)	 &	6.5(2)	& 4378.4	\\
5041.2(1) 	& (3$^-$)  	& 0.95(13)	& 6.34(6)	& 1715.7(1)	 &	1.8(1)	& 3325.4	\\
5348.6(2) 	& (2$^+$)  	& $<$0.1	& 		& 2023.1(2)	 &	0.18(2) & 3325.4	\\
	 	& 	  	& 		& 		& 5348.7(2)	 &	0.38(3) & 0.0  	 	\\
6022.0(2) 	& (3$^-$)  	& 2.76(21)	& 5.70(4)	& 673.4(1)	 &	0.53(2) & 5348.6	\\
	 	& 	  	& 		& 		& 1503.1(1)	 &	0.79(2) & 4518.7	\\
	 	& 	  	& 		& 		& 2696.5(1)	 &	3.2(1)  & 3325.4	\\
6226.7(2) 	& (4$^-$)  	& 0.79(7)	& 6.21(4)	& 1185.8(1)	 &       0.25(2) & 5041.2	\\
	 	& 	  	& 		& 		& 1257.5(1)	 &       0.04(1) & 4969.3	\\
	 	& 	  	& 		& 		& 1848.0(1)	 &       0.87(5) & 4378.4	\\
	 	& 	  	& 		& 		& 1972.7(1)	 &       0.14(2) & 4254.5	\\

\hline
\hline
\end{tabular}
\end{table}

%the logft is the motivation for p2n, intensity balance, half-lives

\begin{table}[!h]
\caption{\label{Tab1} Absolute $\gamma$-ray intensities and neutron emission probabilities in the $^{34}$Mg decay chain. \\}
\begin{tabular}{ccrr}

\hline 
\hline 
\rule{0pt}{3ex}

Nucleus			&$P_n$ 			& $E_{\gamma}$ 		& $I_{abs}$ 									\tabularnewline
						&	[\%]				&	[keV]							&	[\%]												\\
\hline   

\rule{0pt}{3ex}            
$^{34}$Mg			&21(7)			& 								&							                        \tabularnewline
\rule{0pt}{3ex}            
$^{34}$Al			&11(4)($1^+$)& 364.5						&	11.2(15)				                        \tabularnewline
\rule{0pt}{3ex}
$^{34}$Si			&				& 3325.4						&	22(3)					                        \tabularnewline
\rule{0pt}{3ex}
$^{34}$P			&				& 429.0						&	27(3)					                        \tabularnewline
\rule{0pt}{3ex}
$^{34}$S			&				& 2127.2						&	21(2)					                        \tabularnewline
\rule{0pt}{3ex}
$^{33}$Al			&8.5(7) \cite{reeder}	& 1618.0			&	13(5)					                        \tabularnewline
\rule{0pt}{3ex}
$^{33}$Si			&				& 1010.2						&	4.4(16)				                        \tabularnewline
\rule{0pt}{3ex}
$^{33}$P			&				& 1847.8						&	79(17)					                        \tabularnewline
\rule{0pt}{3ex}
$^{32}$Si			&				& 1941.7						&	22(5)					                        \tabularnewline
                           
\hline                       
 
\end{tabular}
\end{table}

\section{Discussion}

\subsection{Shell-model calculations using the SDPF-U-MIX interaction}

Shell model calculations were performed in order to describe the $\beta$ decay of the $4^-$ and $1^+$ states of $^{34}$Al towards excited states in $^{34}$Si using the  {\sc sdpf-u-mix} effective interaction. Similar calculations were performed in the case of $^{34}$Mg decaying into $^{34}$Al \cite{Lica2017}. 

The number of valence particles used in the calculations is 14 neutrons and 6 protons, outside of the $^{16}$O core, in the $sd-pf$ space for the neutrons and in the $sd$ space for the protons. The $2p-2h$ excitations are achieved by promoting a pair of neutrons from the positive $sd$ shell into the negative $fp$ shell. Intuitively, the wavefunction configuration of the lowest negative parity states should be dominated by $1p-1h$ excitations, with the $3p-3h$ only at higher excitation energy. Therefore there is not so much choice, and their clear identification will provide valuable input for shell model calculations. For the positive parity states, the situation is more complex, their configurations being dominated either by the proton $1p-1h$, proton $2p-2h$, as well as neutron $2p-2h$ configurations. 

The {\sc sdpf-u-mix} interaction has two known shortcomings. The first one is that it predicts the positive parity states too low in energy, being an inherited weakness of the USD interaction, which is partly solved if the USDA or USDB interactions are used instead. The second shortcoming is that it shifts up the negative parity states by about 1 MeV compared to the experimental counterparts \cite{himpe}. One way of solving this issue would be through a reduction of the $sd-pf$ monopole gap, but would afterwards affect the positive parity intruder states of $2p-2h$ nature which would appear very low in the spectrum.

The calculations shown in Fig.~\ref{fig:levelScheme} are able to describe well the excitation energy and configuration of the 'normal' $0_1^+$, and intruder $2_1^+$ and $0_2^+$ states.  This has been very well investigated in the past, both experimentally and theoretically \cite{caurier1998, nummela, rotaru} using the {\sc sdpf-nr} and {\sc sdpf-u-si} effective interactions. It should also be noted that in the calculations, the lowest negative parity state has $J^\pi=4^-$, and not $J^\pi=3^-$ as previously determined experimentally\,\cite{nummela}. The present experimental results offer an even better testing ground than ever achieved, especially for the {\sc sdpf-u-mix} interaction which is able to treat higher order intruder configurations and therefore able to predict more precisely the placement and $\beta$-decay strength towards higher excited negative parity states. 

The calculated occupation numbers on each valence orbital corresponding to the $0_2^+$ and $2_{1,2,3}^+$  excited states in $^{34}$Si are shown in Table \ref{tab:wavefunc}.

\begin{table}
\centering
\caption{\label{tab:wavefunc} Neutron ($pf$) and proton ($sd$) occupation numbers on each valence orbital for the first positive parity states in $^{34}$Si calculated using {\sc sdpf-u-mix}. \\ }

\begin{tabular}{c|c|c|c|c|c|c}

\hline 
\hline 
\rule{0pt}{3ex}
\rule{0pt}{3ex}
$J^\pi$ & $1f_{7/2}$ & $2p_{3/2}$ & $2p_{1/2}$ & $1f_{5/2}$ & $1s_{1/2}$ & $1d_{3/2}$ \\
\hline
\rule{0pt}{3ex}
%$0_1^+$	  &	0.38	&	0.02	& 0.0	& 0.01	& 0.0	&	0.0 	\\ 
$0_2^+$	  &	1.82	&	0.13	& 0.02	& 0.12	& 0.75 	&	0.55 \\ 
$2_1^+$	  &	1.73	&	0.12	& 0.02	& 0.09	& 0.86 	&	0.53 \\
$2_2^+$	  &	0.3	&	0.03	& 0.01	& 0.02	& 0.99 	&	0.22 \\
$2_3^+$	  &	1.27	&	0.71	& 0.05	& 0.08	& 0.77 	&	0.50 \\
 \hline

\end{tabular}
\end{table}

\subsection{ Positive parity states in $^{34}$Si }

Allowed GT selection rules in the decay chain of $^{34}$Mg favor the feeding of $0^+, 1^+$ and $2^+$ states in the  $^{34}$Si grand-daughter nucleus. Apart from the previously known $0_1^+$, $0_2^+$ and $2_1^+$ levels, five new $2^+$ states and one $0^+$ state are tentatively proposed  in $^{34}$Si, at 4518.7, 5348.6, 5772.8, 6675.9, 6764.3, and 7475.8 keV, respectively.  The decay pattern of the first four $2^+$ states is similar: a branch towards the $0^+$ ground state and another one to the $2_1^+$ 3325.4-keV level. However, the significant variation of the log($ft$) values, as shown in Fig.~\ref{fig:levelScheme} and Tables 1 and 2, suggests strong differences in the structure among the newly proposed $2^+$ states. The  $0^+$ assignment for the 7475.8-keV level is based on its low log($ft$)  value of 4.72(4) that indicates an allowed transition from the 1$^+$ isomer of $^{34}$Al, as well as its uniquely observed $\gamma$-ray decay pattern to the four $2^+$ states and not to any $0^+$ state. 

The configuration of the positive parity states is rather complex, being dominated either by the proton $1p-1h$, proton $2p-2h$, and neutron $2p-2h$ excitations. 
The calculations using {\sc sdpf-u-mix} for the positive parity states predict the $0_1^+$ ground state to be dominated, at 81\%, by the spherical closed $N=20$ $Z=14$ configuration.  
The 2.58\,MeV $0_2^+$ state is mainly oblate deformed, being dominated by  $2p-2h$ neutrons excitations. 
The 3.45\,MeV $2_1^+$ state belongs to the band of the $0_2^+$ state, being deformed as well. The 4.46\,MeV $2_2^+$ state is of $0p-0h$ nature (spherical), dominated by the $\pi (1d_{3/2})^{-1} (1s_{1/2})^{1}$ configuration with $N=20$ closed.  Therefore, its most probable experimental counterpart is the 5.348 MeV state, whose spin assignment was proposed above. Moreover, the weak $<0.1\%$ $\beta$ feeding from the 1$^+$ intruder ($2p-1h$) isomer of $^{34}$Al suggests that the wavefunction configuration of this $2^+$ state is significantly different, most probably spherical ($0p-0h$), and dominated by the proton excitations inside the $sd$ shell. This state was also populated in the charge-exchange reaction from $^{34}$P \cite{zegers}, which reinforces the present assumption on its proton nature. 
The 5.16\,MeV $2_3^+$, whose experimental counterpart is the $2_2^+$ state at 4.518 MeV, is calculated to be deformed, being the head of the $\gamma$ band. The earlier $0^+$ assignment for the 7475.8-keV level is supported by shell-model calculations, which predict a  $0^+$  state at 7630 keV having a similar log($ft$)  value of 4.69.

\subsection{Negative parity states in $^{34}$Si}

The $(3, 4, 5)^{-}$ states of $^{34}$Si are fed through allowed GT transitions from the decay of  the 4$^-$ ground state of $^{34}$Al, of which only the 7 most intense transitions were previously reported in Ref. \cite{rlica}.  Added to the  information of the log($ft$)  value, the choice of spin assignment for each populated level, among these three possible values, is mostly based on $\gamma$ decay branches. In particular, the tentative $4^-$ assignment of the 4378.4-keV state is based on the fact that it decays solely to the $3^-$ state. Moreover, a  $4^-$ state is predicted close in energy to the $3^-$ level by the present shell model calculations. In absence of octupole collectivity, there is no reason to favour the coupling to $3^-$  over the coupling to $4^-$.  The $5^-$ assignment is deduced from the fact that, despite its high excitation energy, it decays uniquely to the $4^-$ state. The $3^-$ assignment for the 5041.2-keV state is derived from the fact that it decays to the $2^+$ state through an E1 transition, rather than to other negative parity states through M1 or E2 transitions. 

As seen in Fig. \ref{fig:levelScheme}, all the calculated negative parity states below 7\,MeV have their experimental counterparts, demonstrating both the predicting power of the {\sc sdpf-u-mix} interaction and the sensitivity of the present experiment. The calculated negative parity states in $^{34}$Si are however globally shifted up by about 1 MeV as compared to the experimental counterparts, as it was observed also for the case for $^{30,32}$Mg \cite{himpe}.  This probably comes from multiple factors such as an overestimation of the calculated $N=20$ gap ($\sim$5\,MeV, discussed in the following section) or the correlation strength in the $^{34}$Si ground state, both shifting the negative parity states upwards. %Therefore, their clear identification (and in particular that of the unmixed $5^-$ state) provides valuable input for shell model calculations on the size of the neutron shell gap and the amount of correlations induced by these excitations. 
All calculated $0^-,1^-,2^-$ states are above the neutron separation energy and cannot be identified experimentally using the present setup.

\subsection{The $N=20$ shell gap}

In a simplified modeling, the wavefunction configuration of the lowest negative parity states should be dominated by neutron $1p-1h$ excitations across the $N=20$ shell gap. Configurations such as $\nu (f_{7/2})^{1} \otimes \nu (d_{3/2})^{-1} $) or $\nu (p_{3/2})^{1} \otimes \nu (d_{3/2})^{-1} $ lead to states of spin and parity $J^\pi=2^- - 5^-$ or $J^\pi=0^- - 3^-$, respectively. States having wavefunction configurations dominated by $3p-3h$ excitations will have a significantly higher excitation energy. The size of the $N=20$ shell gap is therefore closely linked to the  energy of the unique spin parity $4^-$ and $5^-$ states, most probably of unmixed configuration. The experimental energy of the $5^-$ state in $^{34}$Si (4969 keV) is comparable to other magic nuclei such as $^{36}$S (5206 keV) and $^{40}$Ca (4491 keV)

\begin{figure}[htbp!]
\centering

\includegraphics[width=0.45\textwidth]{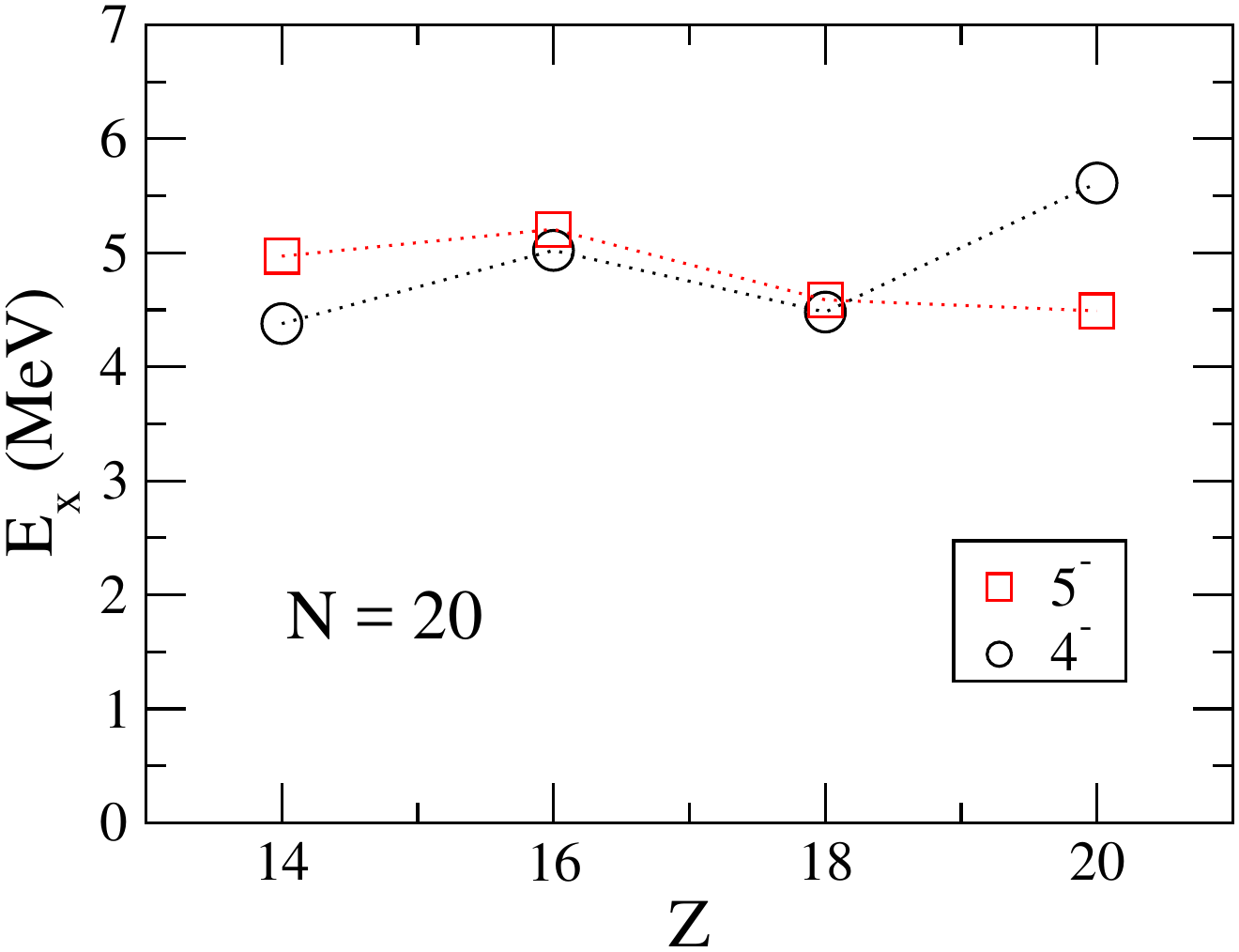}    

\caption{\label{fig:systematics} (Color Online) Excitation energy systematics of $4^-$ and $5^-$ states in $N=20$ even-even nuclei. }
\end{figure}

Fig.\,\ref{fig:systematics} shows the systematics of experimentally known $4^-$ and $5^-$ states in $N=20$ even-even nuclei. It provides an insight towards the evolution of the $sd-pf$ shell gap, provided that these high-spin negative-parity states are dominated by the $\nu (f_{7/2})^{1} \otimes \nu (d_{3/2})^{-1} $ coupling. However, other factors such as correlations have to be taken into account. From the graph can be concluded that the size of $N=20$ gap does not change significantly when moving from $Z=20$ to $Z=14$.

%Start figure at Y axis zero.

In order to estimate the mixing effect and extract the $sd-pf$ shell-gap, calculations using {\sc sdpf-u-mix}  were performed separately for the 0$^+$ state in the $0p-0h$ space, the 0$^+$ state in the $2p-2h$ space and the 4$^-$ state in the $1p-1h$ space. The resulting excitation energies were 0.0, 2.16 and 4.88 MeV, respectively. In the case of the 4$^-$ state, the agreement with the experimental value of 4.4 MeV is reasonable. The full calculation, which includes mixing effects, yields 0.0, 2.58 and 5.31 MeV, respectively. 
The effect of the mixing can be determined by subtracting the calculated excitation energies of the 4$^-$  states in the  $1p-1h$ and full space, respectively, yielding 420~keV. Therefore, by removing 420~keV from the experimental excitation energy of the 4$^-$ state, we can estimate a more realistic value of the shell gap of $\sim$4\,MeV. This confirms that $^{34}$Si can be viewed as a magic nucleus with a large $N=20$ shell gap, however the value obtained is significantly smaller than the $\sim$5\,MeV correlated  $sd-pf$ shell-gap of {\sc sdpf-u-mix}.

\subsection{Mixing between the $0^+_1$ and $0^+_2$ states in $^{34}$Si}

Pure configurations are rarely encountered in atomic nuclei, even at doubly closed shells. In particular, the $0^+$ states in $^{34}$Si (and more generally in all $N=20$ isotones) are likely composed by admixtures of several components that induce their energy shift and the mixing of their wavefunction. The main result of the latter effect is that normally forbidden transitions can increase in strength, while allowed transitions are reduced.  The experimental determination of reduced transition probabilities connecting these 0$^+$ states therefore helps in determining their amount of mixing.  

\begin{figure}[!h]
\begin{minipage}[t]{\linewidth}
    \centering

    \includegraphics[width=8.0 cm]{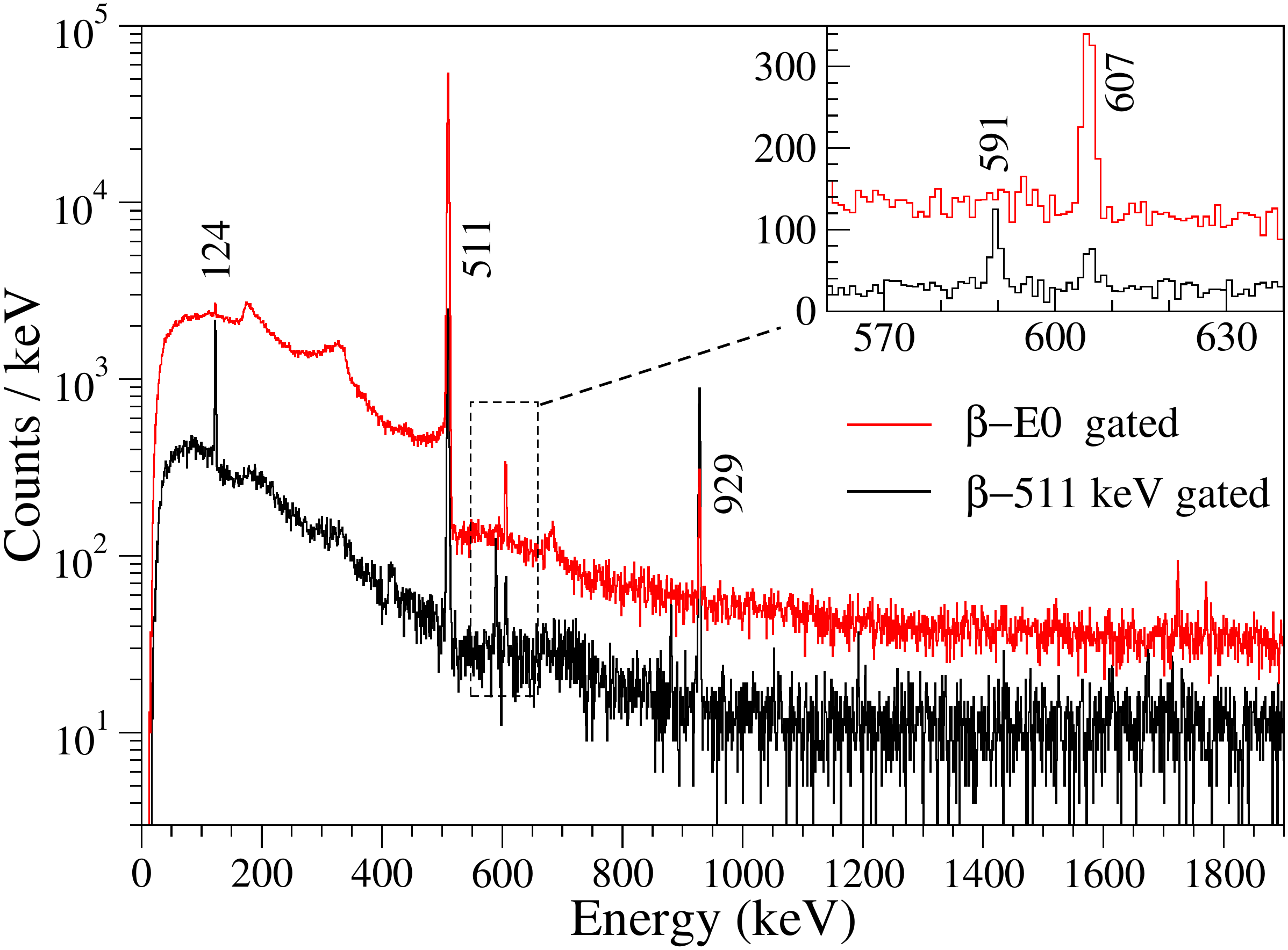}
   
\caption{\label{fig:607keV} (Color Online) Energy spectra of HPGe detectors from the decay of $^{34}$Al in coincidence with double $\beta$ - E0 events in the plastic scintillator (red) [or a single $\beta$ event in the plastic scintillator and a 511-keV $\gamma$-ray in the HPGe detectors (partly originating from the E0 positrons).] The inset shows the region of interest where the 606.8-keV transition can be clearly identified. }
\end{minipage}        
\end{figure}  

The present experiment provides a more precise value for the reduced transition probability $0_2^+ \rightarrow 2_1^+$, as compared to the one earlier reported in \cite{rotaru}, owing to a much more precise determination of the branching ratio ($B_r$) of the two decaying transitions from the $2_1^+$ level, 3325.4 keV to the $0_1^+$ ground state, and 606.8 keV to the $0_2^+$ state.
As first reported in \cite{rotaru}, the weak $E2(2^+\rightarrow0^+_2$) 606.8-keV transition can be observed in the $\beta$ decay of the 4$^-$ ground state of $^{34}$Al in coincidence with the $e^-e^+$ IPF events originating from the 2718 keV $E0(0_2^+\rightarrow0_1^+)$ transition.
About 700 events were collected from the decay of double-hit events ($\beta$ -- E0) in coincidence with 606.8-keV $\gamma$-rays (see Fig.\,\ref{fig:607keV}).  The efficiency for detecting a double-hit event in the plastic scintillator of 57(5)\% was estimated as the squared efficiency for detecting a single event (90(5)\%) corrected by a 70\% factor representing the relative number of decays of the 0$_2^+$ state within the 10 -- 180\,ns integration period.
  Considering the estimated double-hit efficiency and the absolute $\gamma$-ray efficiencies at 606.8 keV and 3325.4 keV, a value of $B_r = 1779(182)$  is obtained, as compared to the previous value of 1380(717) \cite{rotaru}. The time difference distribution between two consecutive hits, shown in Fig.\,\ref{fig:34Mg_fit} allowed for the 0$_2^+$ state half-life measurement, T$_{1/2}$=19.4(5) ns, in perfect agreement with the one previously measured, T$_{1/2}$=19.4(7)\,ns \cite{rotaru}.

\begin{figure}[htbp!]
\centering

\includegraphics[width=0.45\textwidth]{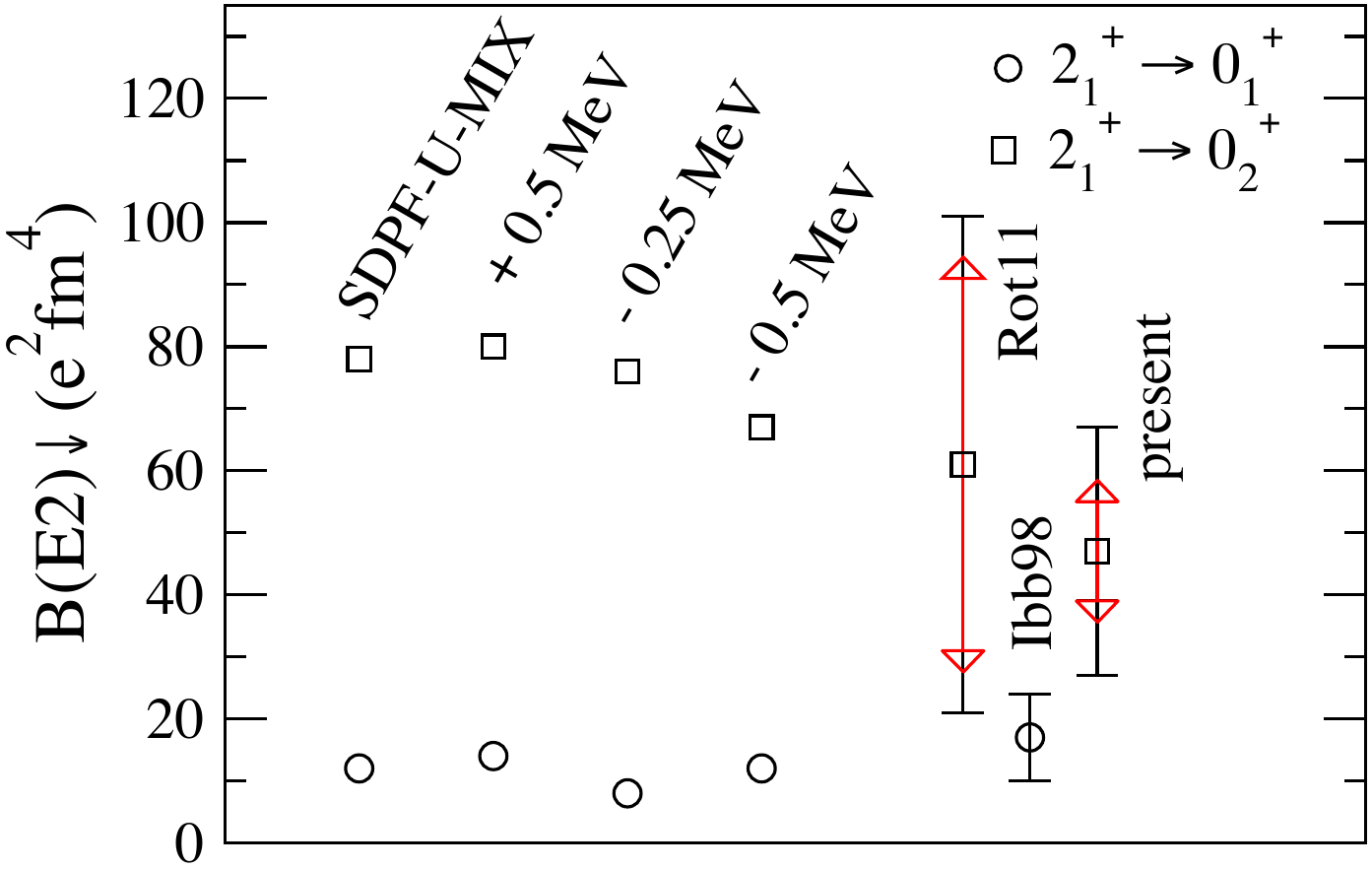}    

\caption{\label{fig:be2} (Color Online) Comparison between shell-model calculations using different offsets (+0.5, -0.25 and -0.5 MeV) for the $N=20$ shell-gap of the SDPF-U-MIX interaction and experimental values of the $B(E2; 2_1^+ \rightarrow 0_1^+)$ and $B(E2; 2_1^+ \rightarrow 0_2^+)$. The previous experimental values were taken from  \cite{ibbotson, rotaru}. The error bars indicated with red arrows represent only the branching ratio uncertainty. }
\end{figure}

Using the presently measured branching ratio, the $B(E2; 0_1^+ \rightarrow 2_1^+) = 85(33)$\,e$^2$fm$^4$ value measured by Coulomb excitation \cite{ibbotson}, and the $(E_{\gamma})^5$ scaling factor for these two E2 606.8 and 3325.5-keV transitions, one obtains  $ B(E2; 2_1^+ \rightarrow 0_2^+)  =  47(19)$\,e$^2$fm$^4$  or $7.2(31)$\,W.u., as compared to the previous value of 61(40)\,e$^2$fm$^4$. A comparison between previous experimental results and shell-model calculations using  {\sc sdpf-u-mix} is shown in Fig.\,\ref{fig:be2}. 

The present 40\% relative uncertainty results from the 39\% uncertainty of the $B(E2; 0_1^+ \rightarrow 2_1^+)$  and the 10\% uncertainty for $B_r$ value. 
Significant improvement for the uncertainties of the experimental values can be achieved only by remeasuring the $B(E2; 0_1^+ \rightarrow 2_1^+)$ value extracted from the Coulomb excitation study \cite{ibbotson} with increased precision and accuracy. 
Furthermore, the $2_3^+$ state at 5348.6 keV will likely be populated by the Coulomb excitation and will decay by 37\% to the $2_1^+$ state through a $\gamma$-ray transition of 2023 keV, and by 63\% to the ground state. An indication of a peak at around 2 MeV can be already observed in the experimental spectrum of $^{34}$Si shown in Fig.\,1 of Ref.\,\cite{ibbotson}. If not singled-out, this contribution will artificially increase the $B(E2; 0_1^+ \rightarrow 2_1^+) \uparrow$.  It is also important to note that this spectrum also contains the contribution of the 1010 keV line from $^{33}$Si, meaning that pure Coulomb excitation was not taken into account to determine the $B(E2; 0_1^+ \rightarrow 2_1^+) \uparrow$ value.

The calculations predict values of $ B(E2; 2_1^+ \rightarrow 0_2^+)  =  78$\,e$^2$fm$^4$ and $ B(E2; 2_1^+ \rightarrow 0_1^+)  =  12$\,e$^2$fm$^4$ which correspond to a mixing between $(0p-0h)$:$(2p-2h)$:$(4p-4h)$ configurations of 90:10:0 for the $0_1^+$ state, 4:89:7 for the $0_2^+$ state and 9:86:5 for the $2_1^+$ state. In order to understand the discrepancy between the measured and calculated $B(E2)$ values above, the $sd-pf$ shell-gap was modified. The discrepancy decreased slightly, within 1$\sigma$, when the gap was decreased by 0.5\,MeV (see Fig.\,\ref{fig:be2}), but this dramatic change in the gap is unreasonable because it would alter the spectroscopic agreement. The calculated $B(E2)$ values have proven to be relatively rigid when modifying the shell-gap, which shows once more that $^{34}$Si behaves as a doubly magic nucleus. 

%Table\,\ref{tab:mixing} shows how the different gap changes ($\Delta$E$_{gap}$) influence the structure of the two $0^+$ states involved and that of the $2_1^+$ state as well. The $B(E2)$ values will depend not only on the amount of $0p-0h$ configurations in the $0_1^+$ state and $2p-2h$ configurations in the $0_2^+$ state, but also by the $4p-4h$ contributions and the configuration mixing of the $2_1^+$ state. It implies that the two-level mixing model employed in Ref.\,\cite{rotaru} does not provide realistic results. 

%In the case of $\Delta$E$_{gap}=-0.25\,MeV$, the B(E2) values are different even if the amount of the $np-nh$ percentages of the three states involved changed very slightly. This can be explained by an internal change of the $np-nh$ components structure of the.

%Fig.\,\ref{fig:systematics2} shows the excitation energy of spherical ($0p-0h$) $2^+$ states in $N=20$ even-even nuclei together with the proposed 5.3486-MeV $2_3^+$ state in $^{34}$Si. It follows the trend set by the $2_1^+$ states in $^{36}$S and $^{38}$Ar. In the case of the doubly-magic $^{40}$Ca, the proton $sd$ is filled and therefore a higher excitation energy is observed.  

%\begin{figure}[htbp!]
%\centering

%\includegraphics[width=0.45\textwidth]{energy_2_plus.pdf}    

%\caption{\label{fig:systematics2} Systematics of spherical ($0p-0h$) $2^+$ states excitation energy in $N=20$ even-even nuclei. }
%\end{figure}

\subsection{Triaxiality in $^{34}$Si?}

A recent study \cite{Han2017} claimed evidence of triaxiality in $^{34}$Si based on the decay pattern of the $2_2^+$ state at 4519 keV to the $0_2^+$ level by a 1800-keV $\gamma$-ray transition, in addition to its already known decay to the $0_1^+$ and $2_1^+$ states. Unfortunately, the authors do not present in their work any experimental spectrum showing the 1800-keV $\gamma$-ray transition. 
The triaxiality claim was also supported by calculations of the $B(E2)$ ratios of the $2_2^+ \rightarrow 2_1^+$ versus $2_2^+ \rightarrow 0_2^+$ transitions from Fig.~6 of Ref.~\cite{Han2017} that amount to 110 for the mean field calculations using the Gogny interaction and 260 for the shell model using the SDPF-M interaction. Assuming pure E2 transitions, and using the experimental energies of the 1193 and 1800-keV transitions, branching ratios of $B_r(1193/1800)$= $I_\gamma(E2; 2_2^+ \rightarrow 2_1^+) / I_\gamma(E2; 2_2^+ \rightarrow 0_2^+)$\,= 14 (Gogny) and 33 (SDPF-M) are found. 

\begin{figure}[htbp!]
\centering

\includegraphics[width=0.48\textwidth]{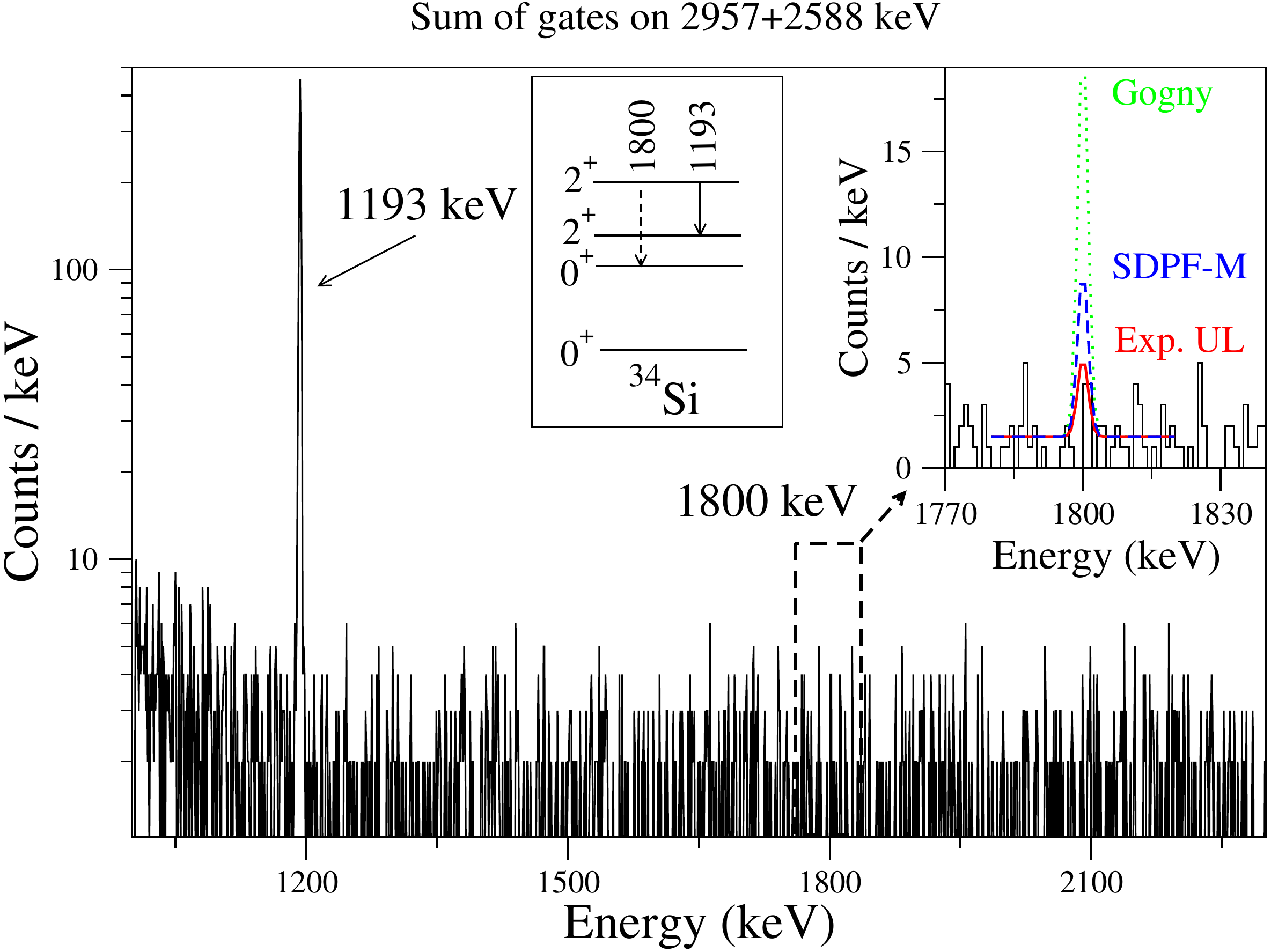}    

\caption{\label{fig:1800kev} (Color Online) Background-subtracted, $\beta$-gated $\gamma$-ray spectra in coincidence with the 2957 ($(0^+) \rightarrow (2^+)$) or 2588-keV ($ \rightarrow (2^+)$) $\gamma$ rays feeding the $(2^+)$ 4518 keV level in $^{34}$Si obtained from the decay of $^{34}$Mg. The left inset shows a simplified level scheme of $^{34}$Si containing the levels of interest and the 1193-keV and tentative 1800-keV transitions. The right inset shows the 1800-keV region and the experimental upper limit (UL) for the intensity of the 1800-keV transition based on a $3\sigma$ confidence level (continuous red line), in comparison to estimations using the SDPF-M (dashed blue line) and Gogny (dotted green line) models \cite{Han2017}.  }
\end{figure}

In the present experimental spectrum shown in Fig.\,\ref{fig:1800kev}, obtained by adding spectra in coincidence with the 2957 ($(0^+) \rightarrow (2^+)$) or 2588-keV ($ \rightarrow (2^+)$) $\gamma$ rays feeding the $(2^+)$ 4518-keV state, a total of 1.8$\times10^5$\,counts are detected in the 1193\,keV $\gamma$-ray peak, however there is no indication of a 1800-keV $\gamma$-ray transition. A similar conclusion is obtained from the $\gamma$-ray spectrum in coincidence with double-hit events from the decay of  $^{34}$Al shown in Fig.\,\ref{fig:607keV}. Therefore we can only extract a lower limit of $B_r(1193/1800)>70$ at a 3$\sigma$ confidence level, which is significantly larger than the calculations. The discrepancy between the calculated and experimental lower limit of $B_r(1193/1800)$ suggests that the claim of triaxiality made in Ref.\,\cite{Han2017} has uncertain experimental grounds.

%Summed gates on 2957 and 2588, background subtracted:
%Area of 1193 = 1117(49) counts
%Area of 1800 = 6(4) counts
%Efficiency at 1193 = 4.2%
%Efficiency at 1800 = 3.2%
%Detection limit = 3 * uncertainty =  12
%Intensity ratio limit = (1117/4.2) / (12/3.2) = > 71

%Beta-E0 gate:
%Area of 607 = 602(54) counts
%Area of 1800 = 48(23) counts
%Efficiency at 607 = 6.1%
%Efficiency at 1800 = 3.2%
%Detection limit = 3 * uncertainty =  69
%Intensity ratio limit (I_607/I_1800) = (602/6.1) / (69/3.2) = > 4.6
%Abs intensity of 1800 =  0.056 / 4.6 = < 0.012
%Abs intensity of 1193 = 0.83
%Intensity ratio limit = 0.83/0.012 = > 69
 
%Figure calculations:
%gaussian function: f(x) = a * exp ( -(x-b)^2/(2c^2) )
%FWHM = 2.7 keV = 2.3548 *c => c = 1.1465 => c^2 = 1.31
%Area = a * sqrt(2*pi *c^2) = a * 3.19
%Area limit of 1800 = 12 counts => a = 12/3.19 = 3.76 
%    -> 3.76 * exp(-($t-1800)^2/2.62) + 2 (background)
%Area of 1800 for GOGNY = 12*70/14 counts => a = 12/3.19 * 70/14= 18.8 
%Area of 1800 for SDPF-M = 12*70/33 counts => a = 12/3.19 * 70/33= 7.97

A different approach when dealing with nuclei showing triaxial deformation employs the $\beta$ and $\gamma$ deformation parameters, which are non-observable quantities used mostly in mean field calculations. 
One of the most popular methods of extracting the $\gamma$ parameter from the experimental observables was defined in Ref.\,\cite{Dav1958}. It relates $\gamma$ to the ratio:
$$ \frac{B(E2) (2^+_{\gamma} \rightarrow  2^+_{yrast})}{B(E2) (2^+_{\gamma}  \rightarrow  0^+_{yrast})} $$ 

As indicated in Table \ref{tab:states_calc}, these states correspond to the 2$^+_3$, the 2$^+_1$ and the 0$^+_2$ states, respectively. We are interested in the shape of the nucleus and not in the shape of the nuclear charge, therefore the ratio value is 38 by considering the mass $B(E2)$'s, corresponding to $\gamma$=26$^\circ$ (according to Table I in Ref.\,\cite{Dav1958} where $\gamma$=30$^\circ$ in case of triaxial deformation).  Even if the difference is not large enough to exclude rigid triaxiality, it must be noted that Ref.\,\cite{Dav1958} does not provide the correct value for $\gamma$, considering the new information reported in Ref.\,\cite{nowacki}.

\begin{table}
\centering
\caption{\label{tab:states_calc} Properties of the first positive parity excited states in $^{34}$Si calculated using {\sc sdpf-u-mix}.  The excitation energies  ($E_x$), spin and parity ($J^{\pi}$), spectroscopic quadrupole moments ( $Q_s$) and reduced transition probabilities for E2 transitions ($B(E2)$)  for charge (c) and mass (m) are indicated. The levels are grouped in bands: S (spherical), D (deformed), G ($\gamma$ band). The Dufour-Zuker effective charges used are 0.46 and 1.31 and the effective masses are 1.77.\\ }

\begin{tabular}{c|c|c|c|c|c|c|c}

\hline 
\hline
\rule{0pt}{3ex}
Band  & $E_x$  & $J^{\pi}$  & $Q_s$(c) & $Q_s$(m)     & $J_f^{\pi}$  & $B(E2)$(c) & $B(E2)$(m)  			\\
           &  [MeV]    & 		          & [$efm^2$]     &  [$fm^2$]    &      &  [\,e$^2$fm$^4$] &  [$fm^4$]    \\
\hline
\rule{0pt}{3ex}

S	& 0.0		& 0$^+_1$ &				&	 &				&	\\
	& 4.46	& 2$^+_2$ &				&  &				&	\\
	& 5.39 	&	3$^+_1$ & 				& &				&		\\
	& 7.13	&	4$^+_3$ & 				&	 &				&	\\

\hline
\rule{0pt}{3ex}

D	& 2.58	& 0$^+_2$ &				&							&				 &				&	\\
	& 3.45	& 2$^+_1$ &	16.2		&	33.2		&	0$^+_1$		&		12	& 39\\ 
	&				&					 &				&					&	0$^+_2$		&		78	& 408\\
	& 5.25 	&	4$^+_1$ & 	14.2		&	26.0 &	2$^+_1$		&		109	& 564\\		  

\hline
\rule{0pt}{3ex}

G	& 5.16	& 2$^+_3$ &	-14.3	&	-30.9 &	2$^+_1$ 	&		38 & 273	\\ 
	&				&					 &				&	 &	0$^+_2$		&		0.01 & 7.2	\\
	& 6.13	& 3$^+_2$ &	0.04		&	0.15 &	2$^+_3$		&		130	& 672\\
	&				&					 &				&	 &	4$^+_1$		&		38 & 230	\\
	& 6.32 	&	4$^+_2$ &		-14.8	&	-34.4 & 	2$^+_3$		&		24 & 184\\ 
	&				&					 &				&	&	4$^+_1$		&		44 &	251 \\
	&				&					 &				&	&	3$^+_2$		&		3	 &	2 \\

\hline

\end{tabular}
\end{table}

Another way to extract the $\beta$ and $\gamma$ parameters from observables (or from calculations in the laboratory frame) is through the use of quadrupole shape invariants (the so-called Kumar invariants) $< \phi | Q^n | \phi >$  \cite{kumar}, which are higher order moments of the quadrupole operator in a given state $\phi$. In our particular case, the state is $\phi$ = 0$^+_2$.  
The great advantage of the Kumar invariants is that they provide not only the values of $\beta$ and $\gamma$ derived from  $< Q^2 >$ and $< Q^3 >$ but also their widths, which require the calculation of  $< Q^4 >$ and  $< Q^6 >$ \cite{nowacki}. When the $\beta$ parameters are considered always positive, the $\gamma$ interval will be 0$^\circ$ -- 60$^\circ$, instead of  0$^\circ$ -- 30$^\circ$ \cite{Dav1958}.
For the deformed structure of interest, $\beta$(charge)=0.47$\pm$0.08  and  $\beta$(mass)=0.42$\pm$0.07, corresponding to values of $\gamma$=46$^\circ$ and $\gamma$=40$^\circ$, respectively. From the computed variance of $< Q^3 >$, we can extract the 1$\sigma$ interval for the $\gamma$ parameter of 29$^\circ$ -- 60$^\circ$ which do not support the claim of triaxial deformation, given the large fluctuations, as discussed in detail in Ref.\,\cite{nowacki}. 

Finally, the arguments against stable triaxiality are also supported by the occupancies of the 2$^+_1$ and 2$^+_3$ states which should be similar if they would pertain to the same intrinsic state. It can be seen in Table \ref{tab:wavefunc} that the occupation numbers for the $2p_{3/2}$ orbit of 0.12 and 0.71, respectively, are significantly different.

%---------------------------------------------------------------------------------------------------------------------

\section{Conclusions} 

The $\beta$-decays of $^{34}$Mg and $^{34}$Al were studied at ISOLDE, CERN, using the recently developed ISOLDE Decay Station. The level scheme of $^{34}$Si was extended up to the neutron separation energy of 7.5 MeV with 11 new levels and 26 new transitions. $\beta-\gamma$ spectroscopic information was extracted for high-spin negative-parity and low-spin positive-parity levels populated independently from the $\beta$-decaying states in $^{34}$Al having spin-parity assignments $J^\pi = 4^-$ dominated by the normal configuration and $J^\pi = 1^+$ by the intruder configuration.
The size of the $N=20$ shell gap was estimated at around 4~MeV, being closely linked to the energy of the unique $J^\pi = 4^-$ and $5^-$ states identified in $^{34}$Si. 
Furthermore, the level of configuration mixing between the normal 0$_1^+$ and intruder 0$_2^+$ and 2$_1^+$ states was studied thanks to a more precise measurement of the $0_2^+ \rightarrow 2_1^+$ reduced transition probability, as compared to the value previously reported in Ref.~\cite{rotaru}. 
Shell-model calculations using the {\sc sdpf-u-mix} interaction were employed in order to interpret the experimental findings and to investigate the recent claims of triaxiality in $^{34}$Si \cite{Han2017}. The present paper concludes that (1) the present experimental and theoretical results do not support the presence of triaxially deformed structures in $^{34}$Si, being consistent with two estimates, based on different theory approaches, which point to a pronounced $\gamma$-softness, in line with the potential energy surface (PES) calculations discussed in Ref.\cite{Han2017}; (2) more precise Coulomb excitation measurements are required in order to lower the uncertainty for the known $B(E2)$ values and to determine the ones corresponding to the newly identified $2^+$ states in $^{34}$Si.

%---------------------------------------------------------------------------------------------------------------------
\section*{Acknowledgments}
\begin{acknowledgments}
This work was partially supported by a grant of the Romanian National Authority for Scientific Research and Innovation, CNCS-UEFISCDI project number PN-II-RU-TE-2014-4-1455, the Romanian IFA Grant CERN/ISOLDE, by Research Foundation Flanders (FWO-Belgium), by GOA/2015/010 (BOF KU Leuven) and by the Interuniversity Attraction Poles Programme initiated by the Belgian Science Policy Office (BriX network P7/12). 
Support from the U.K. Science and Technology Facilities Council, the European Union Seventh Framework through ENSAR (contract no. 262010), the MINECO (Spain) grants FPA2017-87568-P, FPA2015-64969-P, FPA2014-57196, FPA2015-65035-P, Programme ``Centros de Excelencia Severo Ochoa" SEV-2016-0597, the MEYS project SPIRAL2-CZ,EF16-013/0001679, the National Research, Development and Innovation Fund of Hungary with project 
no. K128947 and by the European Regional Development Fund (Contract No. GINOP-2.3.3-15-2016-00034), the German BMBF under contract 05P18PKCIA (ISOLDE) and ``Verbundprojekt 05P2018" is also acknowledged. I.K. was supported by the National Research, Development and Innovation Office of Hungary (NKFIH), contract number PD 124717.
\end{acknowledgments}

%---------------------------------------------------------------------------------------------------------------------%---------------------------------------------------------------------------------------------------------------------

\bibliographystyle{apsrev4-1}
\bibliography{34Si}

\end{document}